\documentclass[useAMS,usenatbib]{mn2e}
\usepackage{times,graphicx,amsmath,amsfonts,amssymb,aas_macros,epstopdf,ulem}
\usepackage{epsfig}
\usepackage[usenames,dvipsnames]{xcolor}
\usepackage{url}
\usepackage{subfigure}

\newcommand{\mpl}{M_{\rm Pl}}
\newcommand{\rd}{{\rm d}}
\newcommand{\tcb}{\textcolor{blue}}
\newcommand{\tcr}{\textcolor{red}}
\newcommand{\tcg}{\textcolor{ForestGreen}}

\def\eg{{\frenchspacing\it e.g.}}

\def\be{\begin{equation}}
\def\ee{\end{equation}}
\def\ba{\begin{eqnarray}}
\def\ea{\end{eqnarray}}
\def\hmpc{h^{-1}\,{\rm Mpc}}
\def\hkpc{h^{-1}\,{\rm kpc}}
\def\dd{\textrm{d}}

\def\de{\delta}
\def\der{\delta_R}
\newcommand{\ad}[1]{\langle\de^{#1}\rangle}
\newcommand{\av}[1]{\langle{#1}\rangle}
\newcommand{\xa}[1]{\overline{\xi}_{#1}}
\def\frac#1#2{{\textstyle{#1\over #2}}}

\def\simlt{\stackrel{<}{{}_\sim}}
\def\simgt{\stackrel{>}{{}_\sim}}

\title[Hierarchical clustering in chameleon  $f(R)$ gravity]{Hierarchical clustering in chameleon $f(R)$ gravity}
\author[Wojciech A.~Hellwing, Baojiu~Li, Carlos~S.~Frenk, Shaun~Cole]{Wojciech A. Hellwing$^{1,2}$\thanks{E-mail: pchela@icm.edu.pl}, Baojiu~Li$^{1}$\thanks{E-mail: baojiu.li@durham.ac.uk},
Carlos S. Frenk$^{1}$ and Shaun Cole$^{1}$\\
$^{1}$Institute for Computational Cosmology, Department of Physics, Durham University, South Road, Durham DH1 3LE, UK\\
$^{2}$Interdisciplinary Centre for Mathematical and Computational Modeling (ICM), Univesity of Warsaw, ul. Pawi\'nskiego 5a, Warsaw, Poland\\
}
\begin{document}

\date{Accepted XXXX . Received XXXX; in original form XXXX}

\pagerange{\pageref{firstpage}--\pageref{lastpage}} \pubyear{2013}

\maketitle

\label{firstpage}

\begin{abstract}
We use a suite of high resolution state-of-the-art N-body Dark Matter simulations of chameleon $f(R)$ gravity
to study the higher order volume averaged correlation functions $\overline{\xi_n}$ together
with the hierarchical $n$-th order correlation amplitudes $S_n=\overline{\xi}_n/\overline{\xi}_2^{n-1}$
and density distribution functions (PDF). We show that under the non-linear modifications of gravity
the hierarchical scaling of the reduced cumulants is preserved. This is however characterised by significant changes of
both the $\overline{\xi_n}$ and $S_n$'s values and their scale dependence with respect to General Relativity gravity (GR). 
In addition, we measure a significant increase of the non linear $\sigma_8$
parameter reaching $14, 5$ and $0.5\%$ in excess of the GR value for the three flavours of our $f(R)$ models.
We further note that the values of the reduced cumulants up to order $n=9$ are significantly increased
in $f(R)$ gravity for all our models at small scales $R\simlt 30\hmpc$. In contrast the values
of the hierarchical amplitudes, $S_n$'s, are smaller in $f(R)$ indicating that the modified gravity
density distribution functions are deviating from the GR case. Furthermore we find that the redshift evolution of relative deviations of the $f(R)$ hierarchical correlation amplitudes is fastest at high
and moderate redshifts $1\leq z \leq4$. The growth of these deviations significantly slows down in the low redshift
universe. We also compute the PDFs and show that for scales below $\sim 20\hmpc$ they are
significantly shifted in $f(R)$ gravity towards the low densities. Finally we discuss the implications of our theoretical predictions for measurements of the hierarchical clustering in galaxy redshifts surveys, including the important problems of the galaxy biasing and redshifts 
space distortions.

\end{abstract}


\section{Introduction}
\label{sec:intro}

The parameters of the standard model of cosmology -
{\it the Lambda Cold Dark Matter Model} - based on the Einstein's theory of {\it General Relativity}
(hereafter LCDM and GR respectively) have been established to an outstanding precision 
\citep[\eg][]{WMAP9,Planck1,galaxy_s8_2,galaxy_s8_1}. The growing observational evidence
has somehow not been fully matched by an appropriate development of theoretical understanding. Alas we are still
left with the riddles and puzzles of Dark Matter and Dark Energy. While there is not much
doubt in the existence of the former, the latter part of the model which is supposed to account 
for the observed accelerated expansion of the Universe \citep{acceleration1,acceleration2}
has an elusive and not fully understand physical nature. The accelerated expansion of the Universe
is usually accounted for by either assuming an extremely low value of the Einstein cosmological
constant $\Lambda$, or by postulating its value to be zero and invoking the background scalar field
to drive the accelerated expansion \citep[\eg][]{de1,de2,de3,de4,de5}. Both approaches however suffer from the well known
coincidence and fine tuning problems \citep[see, \eg][and references therein]{Carroll2001}. 
However it is also possible to obtain an accelerated universe
by modifying the GR equations that govern the background evolution of the Universe \citep[\eg][]{CDTT2004}. 
In other words by implementing a modified gravity model. Such modifications can be done in many ways. In recent years one of the
possible modifications that gained much attention consists of the class of models called the $f(R)$ gravity. Here
the Einstein-Hilbert action is augmented with an arbitrary, intrinsically non-linear function $f$
whose, argument is the Ricci or curvature scalar $R$ \citep[\eg][]{cddett2005,dt2010,sf2010}. The $f(R)$ gravity models
are very interesting as they have potentially rich physics. Not only can modified action fuel the accelerated
expansion but also due to the propagation of an extra scalar degree of freedom can give rise to a fifth-force or
Newtonian gravity enhancement \citep{Chiba2003,Chiba2007}. This in turn can have potentially interesting effects on galaxy and large-scale
structure formation and matter clustering patterns.

If $f(R)$ gravity is to be a feasible theory describing the observable Universe it must pass local gravity tests. Hence 
the fifth-force it introduces must be suppressed in high density regions, like our Solar System. This is achieved by the appropriate choice of the $f(R)$ function
that leads to the so called {\it chameleon mechanism}. This non-linear process traps the scalar field in high density (high curvature)
regions and constrains the local deviations from the usual GR gravity. The intrinsic non-linear character of the chameleon mechanism
makes all predictions for clustering statistics in a $f(R)$ universe very difficult. As the
degree of non-linearity grows both in the matter density and scalar fields perturbation theory predictions quickly become
inaccurate \citep[\eg][]{LiHell2013}. Hence study of the cosmological implications of a chameleon $f(R)$ gravity calls for 
a use of the high-resolution N-body simulations. We base studies presented in this paper on a recently performed suite
of high resolution state-of-the-art chameleon $f(R)$ N-body simulations conducted with a use of the novel code - the \verb#ECOSMOG# 
\citep{lztk2012}.

The standard model of the formation of large scale structure is based on two conventional assumptions. The first is that
structures grew from an initially tiny Gaussian density fluctuations. The second belongs to the mechanism responsible
for growth of perturbations, which is taken to be gravitational instability. This, supplemented by the cold nature of the
main matter ingredient (the Dark Matter), leads to a hierarchical model of structure formation, where the clustering 
proceeds from small to large scales. For power law spectra, $P\propto k^{n_{\rm s}}$, this is always true, provided $n_{\rm s}>-3$. In the $f(R)$
gravity all ingredients of the structure formation model are the same as in the standard one, except for the non-linear
modifications to local gravity - the fifth force, which must lead to non-trivial modifications of the growth mechanism.

All tests of theories for the origin of the large scale structure of the universe, including the modified gravity, rely on
a comparison of model predictions with measurable quantities, derived from observations. The statistical measures we will
discuss in this paper are the low and high order volume-averaged $n$-point correlation functions (or connected moments)
$\overline{\xi}_n$ of the density field. These estimators have
two clear advantages. First, they can be related to the underlying Dark Matter dynamics \citep{1980Peebles,cic_nbody1,cl_stat_dyn,Juszkiewicz1993}.
Second, they can be measured and extracted from galaxy surveys 
\citep[see \eg][and the references therein]{GaztanagaAPM94,2005MNRAS.364..620G,2004MNRAS.352.1232C,2004MNRAS.351L..44B,2001ApJ...551...48Z,skewness_obs1} 
and N-body simulations \citep[\eg][]{npoint_omega_cdm,cic_ana,cic_nbody2,cosmic_error,npoint-halos, Hellwing_npoint}, with a reasonable degree of fidelity and reproducibility.

The set of $n$-point connected moments constitute a simple, yet elegant and complete description of the statistical properties
of the cosmic density field. One of the fundamental predictions of the classical gravitational instability model is that 
the gravitational evolution of the initially Gaussian density field in an expanding universe generates higher order correlations, $\overline{\xi}_n$ 
with $n>2$, which exhibit so-called {\it hierarchical scaling}. That is the higher-order moments scale with variance as
$\overline{\xi}_n = S_n\overline{\xi}_2^{n-1}$. The $S_n$ numbers are called {\it hierarchical amplitudes} and are weakly monotonic
functions of scale $R$. The hierarchical amplitudes only very weakly depend on $\Omega_{\rm M}$ and $\Omega_{\Lambda}$ (the matter and dark energy
cosmic densities). Moreover galaxy biasing and redshift space distortions do not break down the hierarchical scaling of the higher-order
moments. This behaviour of higher-order clustering statistics was largely confirmed for the standard GR model using both theoretical \citep{Fry1984a,Fry1984b,FG1993,FG1994,grav_inst_pt2,grav_inst_pt4} 
and observational evidence \citep{GaztanagaAPM94,skewness_obs1,2005MNRAS.364..620G,2004MNRAS.352.1232C,2004MNRAS.351L..44B}.

Because the hierarchical scaling and clustering was so thoughtfully tested for standard GR model paradigm it is crucial
to establish predictions for the correlation hierarchy in the $f(R)$ and any other realistic modified gravity model. This is 
the main goal and aim of this paper. The high-order correlations hierarchy was studied for a simple model of a fifth-force
modified gravity by \citet{Hellwing_npoint}. They found that even in the regime of modified dynamic the hierarchical scaling 
is preserved, although the values of $S_n$'s and their scale and time dependence deviate from the standard model. Their model
however assumed very simple phenomenological form of modified gravity. In this work we study for the first time the high-order correlations
for a more physically motivated $f(R)$ gravity model with a full treatment of the non-linear chameleon mechanism.

\section{The \MakeLowercase{$f($}$R)$ gravity theory}
 
\label{sect:fr_gravity}

This section is devoted to a brief review of the $f(R)$ gravity theory and its theoretical properties.

\subsection{The \MakeLowercase{$f$}$(R)$ gravity model}

\label{subsect:fr}

The $f(R)$ gravity model \citep{cddett2005} is a generalisation of GR achieved by replacing the Ricci scalar $R$ in 
the Einstein-Hilbert action with an algebraic function $f(R)$~\citep[see \eg][for most recent reviews]{sf2010, dt2010}
\ba
\label{eq:fr_action}
S &=& \int{\rm d}^4x\sqrt{-g}\left\{\frac{\mpl^2}{2}\left[R+f(R)\right]+\mathcal{L}_{\rm m}\right\},
\ea
in which $\mpl$ is the reduced Planck mass, $\mpl^{-2}=8\pi G$, $G$ is Newton's constant, $g$ the determinant of the metric $g_{\mu\nu}$ and $\mathcal{L}_{\rm m}$
the Lagrangian density for matter and radiation fields (including photons, neutrinos, baryons and cold dark matter). By designing the functional form of $f(R)$ one can fully specify
a $f(R)$ gravity model.

Varying the action, Eq.~(\ref{eq:fr_action}), with respect to the metric field $g_{\mu\nu}$, one obtains the modified Einstein equation
\begin{eqnarray}\label{eq:fr_einstein}
G_{\mu\nu} + f_RR_{\mu\nu} -g_{\mu\nu}\left[\frac{1}{2}f-\Box f_R\right]-\nabla_\mu\nabla_\nu f_R = 8\pi GT^m_{\mu\nu},
\end{eqnarray}
where $G_{\mu\nu}\equiv R_{\mu\nu}-\frac{1}{2}g_{\mu\nu}R$ is the Einstein tensor, $f_R\equiv \rd f/\rd R$, $\nabla_{\mu}$ is the covariant derivative
compatible with $g_{\mu\nu}$, $\Box\equiv\nabla^\alpha\nabla_\alpha$ and $T^m_{\mu\nu}$ is the energy momentum tensor of matter and radiation fields. 
Eq.~(\ref{eq:fr_einstein}) is a fourth-order differential equation, but can also be considered as the standard second-order equation of GR with a new 
dynamical degree of freedom, $f_R$, the equation of motion of which can be obtained by taking the trace of Eq.~(\ref{eq:fr_einstein})
\begin{eqnarray}\label{eq:fr_eom}
\Box f_R = \frac{1}{3}\left(R-f_RR+2f+8\pi G\rho_{\rm m}\right),
\end{eqnarray}
where $\rho_{\rm m}$ is the matter density. The new degree of freedom $f_R$ is often dubbed {\it the  scalaron} in the literature \citep[\eg][]{zlk2011}.

If the background Universe is described by the flat Friedmann-Robertson-Walker (FRW) metric, the line element of the real, perturbed, Universe can be written in the conformal Newtonian gauge as
\begin{eqnarray}
\rd s^2 = a^2(\eta)\left[(1+2\Phi)\rd\eta^2 - (1-2\Psi)\rd x^i\rd x_i\right],
\end{eqnarray}
where $\eta$ and $x^i$ are the conformal time and comoving coordinates, $\Phi(\eta,{\bf x})$ and $\Psi(\eta,{\bf x})$ are respectively the Newtonian
potential and perturbed spatial curvature, which are functions of both time $\eta$ and space ${\bf x}$; $a$ denotes the scale factor of the
Universe with a normalisation of $a=1$ today.

We will be mainly interested in large-scale structure on scales much smaller than the Hubble scale. Since the time variation of $f_R$ is very small in
the models to be considered below, we shall work in the quasi-static limit by neglecting the time derivatives of $f_R$. Under this limit, 
the $f_R$ equation of motion - Eq.~(\ref{eq:fr_eom}), reduces to
\begin{eqnarray}\label{eq:fr_eqn_static}
\vec{\nabla}^2f_R &=& -\frac{1}{3}a^2\left[R(f_R)-\bar{R} + 8\pi G\left(\rho_{\rm m}-\bar{\rho}_{\rm m}\right)\right],
\end{eqnarray}
where $\vec{\nabla}$ is the three dimensional gradient operator (an arrow is used to distinguish this from the $\nabla$ introduced above), and the overbar takes
the background average of a quantity. Note that $R$ can be expressed as a function of $f_R$ by reverting $f_R(R)$.

Similarly, the Poisson equation, which governs the behaviour of the Newtonian potential $\Phi$, simplifies to
\begin{eqnarray}\label{eq:poisson_static}
\vec{\nabla}^2\Phi &=& \frac{16\pi G}{3}a^2\left(\rho_{\rm m}-\bar{\rho}_{\rm m}\right) + \frac{1}{6}a^2\left[R\left(f_R\right)-\bar{R}\right],
\end{eqnarray}
by neglecting terms involving time derivatives of $\Phi$ and $f_R$, and using Eq.~(\ref{eq:fr_eqn_static}) to eliminate $\vec{\nabla}^2f_R$.

The above equations imply two potential cosmological effects of the scalaron field: (i) the background expansion of the Universe can
be modified by the new terms in Eq.~(\ref{eq:fr_einstein}) and (ii) the relationship between the gravitational potential $\Phi$ and the matter density 
field is modified, which can cause changes in the matter clustering and growth of density perturbations. Evidently, when $|f_R|\ll1$, we 
have $R\approx-8\pi G\rho_{\rm m}$ according to Eq.~(\ref{eq:fr_eqn_static}) and so Eq.~(\ref{eq:poisson_static}) reduces to the normal 
Poisson equation of GR; when $|f_R|$ is large, we instead have $|R-\bar{R}|\ll8\pi G|\rho_{\rm m}-\bar{\rho}_{\rm m}|$ and then Eq.~(\ref{eq:poisson_static}) 
reduces to the normal Poisson equation with $G$ rescaled by $4/3$. The value $1/3$ is the maximum enhancement factor of gravity in $f(R)$ models,
independent of the specific functional form of $f(R)$. The choice of $f(R)$, however, is important because it determines the scalaron dynamics 
and therefore when and on which scale the enhancement factor changes from 1 to $4/3$: scales much larger than the range of the modification 
to Newtonian gravity mediated by the scalaron field ({\it i.e.}, the Compton wavelength of $f_R$) are unaffected and gravity is not enhanced there,
while on small scales, depending on the environmental matter density, the $1/3$ enhancement may be fully realised -- this results in a scale-dependent 
modification of gravity and therefore a scale-dependent
growth rate of structures.

\subsection{The chameleon mechanism}
 
\label{subsect:fr_cham}

The local test of gravity, based on the Solar System observations, provide tight constraints on any deviations from a Newtonian gravity (references).
The classical $f(R)$ model is than strongly ruled out due to its factor-of-$4/3$ enhancement to the strength of Newtonian gravity (references).
However, it can be shown that, if $f(R)$ is chosen appropriately \citep{bbh2006,ftbm2007,nv2007,lb2007,hs2007,bbds2008}, the model can exploit
the so-called chameleon mechanism \citep{kw2004,ms2007} to suppress the gravity force enhancement and therefore pass the experimental constraints in high matter density
regions such as our Solar system.

The basic idea of the chameleon mechanism is the following: the modifications to Newtonian gravity can be considered as an extra, or fifth, force
mediated by the scalaron field $f_R$. Because the scalaron itself is massive, this extra force is of the Yukawa type, decaying exponentially 
$\exp(-mr)$, in which $m$ is the scalaron mass, as the distance $r$ between two test masses increases. In high matter density environments, $m$ is very
heavy and the exponential decay causes a strong suppression of the force over distance. In reality, this is equivalent to setting $|f_R|\ll1$ in high 
density regions because $f_R$ is the potential of the fifth force, and this leads to the GR limit as we have discussed above.

Consequently, the functional form of $f(R)$ is crucial in determining whether the fifth force can be sufficiently suppressed in high density
environments. In this paper we will focus on the $f(R)$ Lagrangian proposed by \citet{hs2007}, for which
\begin{eqnarray}\label{eq:hs}
f(R) = -M^2\frac{c_1\left(-R/M^2\right)^n}{c_2\left(-R/M^2\right)^n+1},
\end{eqnarray}
where $M^2\equiv8\pi G\bar{\rho}_{\rm m0}/3=H_0^2\Omega_{\rm M}$, with $H$ being the Hubble expansion rate and $\Omega_{\rm M}$ the present-day fractional
density of matter. Throughout this paper a subscript $0$ always denotes the present-day ($a=1$, $z=0$) value of a quantity. It was shown by \citet{hs2007}
that $|f_{R0}|\lesssim0.1$ is necessary to evade the Solar system constraints but the exact constraint depends on the behaviour of $f_R$ in galaxies as well.

In the background cosmology of this $f(R)$ model, the scalaron field $f_R$ always sits close to the minimum of the effective potential that governs its 
dynamics, defined as
\begin{eqnarray}
V_{\rm eff}\left(f_R\right) \equiv \frac{1}{3}\left(R-f_RR+2f+8\pi G\rho_{\rm m}\right),
\end{eqnarray}
around which it quickly oscillates with small amplitude \citep{bdlw2012}. Therefore we find
\begin{eqnarray}
-\bar{R} \approx 8\pi G\bar{\rho}_{\rm m}-2\bar{f} = 3M^2\left(a^{-3}+\frac{2c_1}{3c_2}\right).
\end{eqnarray}
To match the background evolution of the $\Lambda$CDM model which is tightly constrained nowadays \citep{WMAP9,Planck1}, we set
\begin{eqnarray}
\frac{c_1}{c_2} = 6\frac{\Omega_\Lambda}{\Omega_{\rm M}}
\end{eqnarray}
where $\Omega_{\rm M}$ and $\Omega_\Lambda$ are respectively the present-day fractional energy densities of the dark matter and dark energy.

We adopt standard LCDM model normalisation, by taking  $\Omega_\Lambda=0.76$ and $\Omega_{\rm M}=0.24$
\footnote{These values are used in the $f(R)$ simulations extensively in the literature,
and we use them in the simulations used in this work in order to compare with previous work.}. We find that $|\bar{R}|\approx41M^2\gg M^2$, and
this simplifies the expression of the scalaron to
\begin{eqnarray}
f_R \approx -n\frac{c_1}{c_2^2}\left(\frac{M^2}{-R}\right)^{n+1}.
\end{eqnarray}
Therefore, the two free parameters $n$ and $c_1/c_2^2$ completely specify the Hu-Sawicki $f(R)$ model. Furthermore, $c_1/c_2^2$ is related to the value of 
the scalaron today, $f_{R0}$, by
\begin{eqnarray}
\frac{c_1}{c_2^2} = -\frac{1}{n}\left[3\left(1+4\frac{\Omega_\Lambda}{\Omega_{\rm M}}\right)\right]^{n+1}f_{R0}.
\end{eqnarray}
In the present paper we will study three $f(R)$ models with $n=1$ and $|f_{R0}|=10^{-6}, 10^{-5}, 10^{-4}$, which we refer to as F6, F5 and F4 
respectively. These choices of the value of $|f_{R0}|$ are decided to cover the whole parameter space that would be cosmological interesting: if $|f_{R0}|>10^{-4}$
then the $f(R)$ model violates the cluster abundance constraints \citep{svh2009}, and if $|f_{R0}|<10^{-6}$ then the difference from 
$\Lambda$CDM will be too small to be observable in practice, as we will show later.

\section{The N-body simulations of \MakeLowercase{$f$}$(R)$ gravity}

\label{sec:nbody}
From Eqs.~(\ref{eq:fr_eqn_static}, \ref{eq:poisson_static}) we see that, with the matter density field known, we can solve for the scalaron field $f_R$
using Eq.~(\ref{eq:fr_eqn_static}) and substitute the result into the modified Poisson equation (\ref{eq:poisson_static}) to solve for $\Phi$. Once the $\Phi$ is
obtained, we can differentiate it to get the modified gravitational force which determines how the particles move in space. These are basically
what we need to do in $f(R)$ $N$-body simulations to evolve the matter distribution.

The major challenge in $f(R)$ $N$-body simulations is to solve the scalaron equation of motion, Eq.~(\ref{eq:fr_eqn_static}), which is
highly nonlinear when the chameleon mechanism is at work. One way to do this is to use a mesh (or a set of meshes) on which $f_R$ could be solved using say relaxation 
methods. This implies that mesh-based $N$-body codes are most convenient. On the other hand, tree-based codes are more difficult to apply here, as we do 
not have any analytical formula for the modified force law (such as the $r^{-2}$-law in the Newtonian case) due to the complexities stemming from the breakdown of the superposition
principal.

$N$-body simulations of $f(R)$ gravity and related theories have previously been performed by 
\citet{oyaizu2008,olh2008,sloh2009,zlk2011,lz2009,lz2010,schmidt2009,lb2011,bbdls2011,dlmw2012}. As the strong nonlinearity of Eq.~(\ref{eq:fr_eqn_static})
means that the code spends a significant portion of the computing time on solving it, most of these simulations were
limited by either the box size or resolution, or both. For this work we have run simulations using the recently developed {\sc ecosmog} 
code \citep{lztk2012}. {\sc ecosmog} is a modification of the mesh-based $N$-body code {\sc ramses} \citep{ramses}, which is efficiently parallelised 
using MPI and can therefore better utilise the supercomputing resources and improve on both simulation resolutions and box sizes. More technical details 
of the code can be found in \cite{lztk2012,LiHell2013,jblkz2012} and we will not repeat here.

The simulations used in this work are summarised in Table~\ref{table:simulations}. All of them are described by the same set of cosmological parameters 
so that the background cosmology for all models is the same in practice (the difference caused by the different $f(R)$ model parameters is negligble). 
The values of cosmological parameters for our runs are
the following: $\Omega_{\rm M} = 0.24$, $\Omega_\Lambda = 0.76$, $h = 0.73$, $n_{\rm s} = 0.958$ and $\sigma_8 = 0.80$, where $h\equiv H_0/(100$~km/s/Mpc) is 
the dimensionless Hubble parameter today, $n_{\rm s}$ is
the scalar index of the primordial power spectrum and $\sigma_8$ is the linear rms density fluctuation measured in 
spheres of radius 8$h^{-1}$Mpc at $z=0$. 

\begin{table*}
\caption{Some technical details of the simulations performed for this work. F6, F5 and F4 are the labels of the Hu-Sawicki $f(R)$ models with $n=1$ and
$|f_{R0}|=10^{-6}, 10^{-5}, 10^{-4}$ respectively. Here $N_{\rm p}$ is the total number of N-body particles used and $k_{\rm Nyq}$ denotes the Nyquist frequency.
Two parameters set the resolutions of our simulations, they are the force resolution $\varepsilon$ and the mass resolution $m_{\rm p}$.
The last column lists the number of realisations for each simulation.}
\begin{tabular}{@{}lccccccc}
\hline\hline
models & $L_{\rm box}$ & $N_{\rm p}$ & $k_{\rm Nyq}$ $[h~\textrm{Mpc}^{-1}]$ & $\varepsilon$ [$\hkpc$] & $m_{\rm p}$ [$M_{\odot}/h$] & number of realisations \\
\hline
$\Lambda$CDM, F6, F5, F4 & $1.5h^{-1}$Gpc & $1024^3$ & 2.14 & 22.9 & $2.094\times 10^{10}$ & $6$ \\
$\Lambda$CDM, F6, F5, F4 & $1.0h^{-1}$Gpc & $1024^3$ & 3.21 & 15.26 & $6.204\times 10^{9}$ & $1$ \\
\hline
\end{tabular}
\label{table:simulations}
\end{table*}

All models in each simulation share the same initial condition computed at the initial time of 
$z_{\rm i} = 49$ using the Zel'dovich approximation \citep{za}. Note that in general the modified gravity affects the generation of the initial 
condition too \citep{lb2011}, but in our case here we can use the same initial conditions for all three $f(R)$ models because
the differences in clustering between GR and the $f(R)$ models are negligible at early times (redshifts higher than a few). 
The fact that we use the same initial conditions for all simulations in a given set is an advantage: since the initial density fields for the GR and 
$f(R)$ simulations have the same phases, any difference in the clustering amplitudes that we find at later times will be a direct consequence of the 
different dynamics between the two cosmologies.

\subsection{Density estimation}
\label{subsect:density}

We aim to compute higher-order statistics of the density field. From the computation point of view it is important to reconstruct
high-resolution and high-quality density fields from the DM particles of our simulations. This is crucial for the accuracy of our
later computations, as the high-order moments are strongly affected by shot-noise and resolution effects. We choose to employ 
{\it the Delaunay Tesselation Field Estimator} method (hereafter DTFE) \citep{sv2000,vs2009}. We use the publicly available
software implementing the DTFE method written by \citet{cv2011}.
This approach consist of a natural method of reconstructing a volume-weighted and continuous density field from 
a discrete set of sampling points. The field reconstructed using the DTFE method is largely shot-nose free down 
to the resolution limit (the fluid limit) of the point distribution. The shot-noise is only present due to the intrinsic Monte Carlo
sampling of the density inside the Delaunay cell. To suppress this source of error we use 1000 Monte Carlo sampling points 
for each of the Delaunay tetrahedron. For our purpose we decided to interpolate the DTFE density field over a $1024^3$ regular
sampling mesh. This sets our spatial resolution of $1.46\hmpc$ and $0.97\hmpc$ for the $1500\hmpc$ and $1000\hmpc$ box simulations respectively.
This is equal to the Nyquist scales for this simulations. For any discretely sampled field the fluid limits breakdown close to its
Nyquist scale/frequency. Thus we will limit our analysis to the scales twice the resolution limit (respectively 3 and $2\hmpc$).

\section{Hierarchical clustering}
\label{sec:h_cluster}

\subsection{The definitions}
\label{subsect:definitions}

We start by introducing the dark matter density field, given by the expression
\be
\label{eqn:deltarho}
\rho(\vec{x},t) = \av{\rho(t)}\,[1 + \delta(\vec{x},t)] \, ,
\ee
where $\left<\rho(t)\right>$ is the ensemble average of the dark matter density (the mean background density
of the Universe) at time $t$, and $\delta(\vec{x},t)$ 
(the local density contrast) describes local deviations from homogeneity. For clarity we will drop the explicit time and 
position dependence of the density contrast in most of our equations.
Structure formation is driven only by the spatially fluctuating part of the gravitational potential, $\phi(\vec{x},t)$,
induced by the density fluctuation field $\delta$. In $f(R)$ cosmologies, however, we expect that in regions where
the fifth force is not screened by the chameleon mechanism the standard gravitational
potential will be enhanced by the scalaron as described by Eq. (\ref{eq:poisson_static}). Thus we expect that 
clustering will be enhanced in our $f(R)$ models at small and moderate scales. This was already shown 
for the two-point statistics \citep{matter_pk_FR,LiHell2013}.

\subsubsection{The cumulants of the density field}

The nonlinear gravitational evolution of the density field $\de$ drives the field (and its distribution function)
away from the initial Gaussian distribution. Deviations of a field from Gaussianity can be characterised 
by {\it cumulants} or {\it reduced moments}. Thus the basic objects of our analysis 
are the cumulants  of the density field distribution function $p(\de)$.
The $n$-th cumulant of the distribution function $\de$ is defined by recursive relation to the $n$-th
moments. This relation can be expressed by cumulant generating function \citep[eg.][]{Lokas_kurt}
\be
\label{eqn:gen_func}
\ad{n}\equiv M_n = {\partial^n \ln\av{e^{t\de}}\over \partial t^n}\bigg|_{t=0}\,.
\ee
The cumulants now can be expressed in terms of the central moments, in particular, for the first 9 cumulants we have \citep{ber1994,GaztanagaAPM94}
\ba
\label{eqn:cumulants}
\ad{}_{\rm c} &=& 0,\,\,\textrm{(the mean)}\nonumber\\
\ad{2}_{\rm c} &=& \ad{2}\equiv\sigma^2,\,\,\textrm{(the variance)}\nonumber\\
\ad{3}_{\rm c} &=& \ad{3},\,\,\textrm{(the skewness)}\nonumber\\
\ad{4}_{\rm c} &=& \ad{4} - 3\ad{2}_{\rm c}^2,\,\,\textrm{(the kurtosis)}\nonumber\\
\ad{5}_{\rm c} &=& \ad{5} - 10\ad{3}_{\rm c}\ad{2}_{\rm c},\nonumber\\
\ad{6}_{\rm c} &=& \ad{6} - 15\ad{4}_{\rm c}\ad{2}_{\rm c} - 10\ad{3}_{\rm c}^2+30\ad{2}_{\rm c}^3,\nonumber\\
\ad{7}_{\rm c} &=& \ad{7} - 21\ad{5}_{\rm c}\ad{2}_{\rm c} - 35\ad{4}_{\rm c}\ad{3}_{\rm c} + 210\ad{3}_{\rm c}\ad{2}_{\rm c}^2,\nonumber\\
\ad{8}_{\rm c} &=& \ad{8} - 28\ad{6}_{\rm c}\ad{2}_{\rm c} - 56\ad{5}_{\rm c}\ad{3}_{\rm c} - 35\ad{4}_{\rm c}^2\nonumber\\
         &+& 420\ad{4}_{\rm c}\ad{2}_{\rm c}^2 + 560\ad{3}_{\rm c}^2\ad{2}_{\rm c} - 630\ad{2}_{\rm c}^4,\nonumber\\
\ad{9}_{\rm c} &=& \ad{9} -36\ad{7}_{\rm c}\ad{2}_{\rm c} - 84\ad{6}_{\rm c}\ad{3}_{\rm c} - 126\ad{5}_{\rm c}\ad{4}_{\rm c}\nonumber\\
         &+& 756\ad{5}_{\rm c}\ad{2}_{\rm c}^2 + 2520\ad{4}_{\rm c}\ad{3}_{\rm c}\ad{2}_{\rm c} + 560\ad{3}_{\rm c}^3\nonumber\\
         &-& 7560\ad{3}_{\rm c}^2\ad{2}_{\rm c}^3.
\ea
In general the value of the $n$-th cumulant is the value of the $n$-th moment of the distribution
from which one must subtract the results of all the decompositions of a set of $n$ points in its subsets
multiplied (for each decomposition) by the cumulants corresponding to each subset \citep{ber1994}.

For the Gaussian field with a zero mean all connected moments die out except the variance $\ad{2}$. In the classical
random field theory the first two non-vanishing cumulants after variance have special meaning as they measure particular
shape departures of the distribution function from Gaussianity. The skewness is a measure of the asymmetry of the 
distribution and the value of kurtosis characterise the flattening of the tails with respect to a Gaussian. Higher 
cumulants measure more complicated shape deviations of the distribution function. 

\subsubsection{The hierarchical amplitudes}

It is well established \citep[eg.][]{ber1992,Juszkiewicz1993,BCGS_book,npoint_omega_cdm,cic_nbody1} that gravitational
evolution of the initially Gaussian field creates and preserves quasi-Gaussian clustering hierarchy
of cumulants that is characterised by the {\it hierarchical scaling}
\be
\label{eqn:h-scalling}
\ad{n}_{\rm c} = S_n\ad{2}_{\rm c}^{n-1} = S_n\sigma^{2n-2}\,,
\ee
where the $S_n$ are called {\it hierarchical amplitudes} or {\it reduced cumulants} and for
unsmoothed field are constant. For example for $\Omega=1$ Universe \citep{1980Peebles} 
found the reduced skewness to be $S_3=34/7\cong 4.86$ while \citep{ber1994} estimated the reduced kurtosis 
to be $S_4=60712/1223\cong45.89$.

\subsubsection{The smoothing}

The observational data stemming from recent and future galaxy redshift surveys allow one 
to estimate the cumulants of a {\it smoothed} density field. In order to make any
testable predictions we need to account for that fact. Thus it is handy to define a
new field $\overline{\de}$, whose value at any point $\mathbf{x}$ in space is either the average value of
$\de$ in some defined volume, centred on $\bf{x}$, or an integral over volume, taken with some
weighting function. Therefore we define the smoothed density contrast field as
\be
\label{eqn:de-smooth-real}
\der(\mathbf{x}) \equiv \overline{\de}(\mathbf{x}) = \int \de(\mathbf{x}')W(|\mathbf{x}-\mathbf{x}'|/R_{\rm w})\dd^3x'\,,
\ee
where $W(x/R_{\rm w})$ is a spherically symmetric window or smoothing function. We will consider only filters that
are spherically symmetric with a finite effective half-width $R_{\rm w}$ and in addition are normalised to unity
\be
\label{eqn:filter-norm}
\int W(y)\dd^3y = 1\quad\textrm{with}\quad \int W(y)y^2\dd^3y = R_{\rm w}^2\,.
\ee
Smoothing over a ball of radius $R_{\rm TH}\equiv R_{\rm w}$ is called {\it top-hat} smoothing. We should denote here
that effects of smoothing and gravitational dynamical evolution commute only for $\de^1=\de$, for
second and higher orders, these two processes are not interchangeable \citep[see also][]{smoothing_comute, cl_stat_dyn}.

It is convenient to define now the Fourier representation of the real density field. The Fourier space
image of a density field is
\be
\label{eqn:de_k_def}
\de(\mathbf{k}) \, \equiv \, (2\pi)^{-3/2}\,\int \de(\mathbf{x})\,e^{-i\mathbf{k}\cdot\mathbf{x}}\,d^3x\,\,
\ee
The main advantage of the frequency space is that a convolution of real-space functions from equation (\ref{eqn:de-smooth-real}) is
exchanged by a simple multiplication. In Fourier space the smoothed density contrast is then
\be
\label{eqn:de-smooth-k}
\der(\mathbf{k})= \de(\mathbf{k})W(\mathbf{k}R_{\rm TH})\,,
\ee
where
\be
\label{eqn:top-hat-k}
W(\mathbf{k}R_{\rm TH})=(2\pi)^{-3/2}\,\int W(\mathbf{x}/R_{\rm TH})\,e^{-i\mathbf{k}\cdot\mathbf{x}}\,d^3x\,,
\ee
is the Fourier image of the window function. For the spherical top-hat window that we use the transformation yields $W(\mathbf{k}R_{\rm TH})=(3/kR_{\rm TH})j_1(kR)$
with $j_1$ being a spherical Bessel function of the first kind. Now if we want to obtain a smoothed real space density field $\der$ we can employ
the top-hat filtering in Fourier space and use inverse transformation to get back the top-hat smoothed real space field
\be
\label{eqn:de-k-to-real}
\der(\mathbf{x})=(2\pi)^{-3/2}\,\int \der(\mathbf{k})W(\mathbf{k}R_{\rm TH})\,e^{i\mathbf{k}\cdot\mathbf{x}}\,d^3k\,,
\ee
We will use this technique extensively in our studies as it is very efficient computationally. Finally we can define
{\it volume-averaged} $n-$point correlation function of the field $\der$ as
\ba
\label{eqn:xi-vol-def}
\overline{\xi}_n(R_{\rm TH})\equiv\av{\der^n}_{\rm c} =\\\nonumber
 \int \dd^3 x_1\ldots\dd^3 x_n\xi(\mathbf{x_1}\ldots\mathbf{x_n})W(x_1/R_{\rm TH})\ldots W(x_n/R_{\rm TH})\,.
\ea

The effects of smoothing on hierarchical amplitudes and density cumulants were studied within a perturbation theory (hereafter PT) framework by \cite{ber1994} and \cite{Juszkiewicz1993}. 
They both found that smoothing induces weak scale dependence of the $S_n$'s and this effect is quantified by various combinations (depending on the cumulant order)
of the logarithmic slope of the variance $\gamma_n$, which is defined as
\be
\label{eqn:pt_gamma}
\gamma_n(R_{\rm TH})\equiv {\dd^n\textrm{log}\sigma^2(R_{\rm TH})\over \dd\textrm{log}^n R_{\rm TH}}\,.
\ee
For a smoothed reduced skewenss $S_3$ and kurtosis $S_4$ the PT predicts \citep{ber1994}
\ba
\label{eqn:s3s4_gamma}
S_3 = {34\over 7} + \gamma_1\,,\nonumber\\
S_4 = {60712\over 1323} + {62\over 3}\gamma_1 + {7\over 3}\gamma_1^2+{2\over 3}\gamma_2\,.
\ea
For a density field characetrised by the spectral index $-3\leq n\leq1$ the values of the logarithmic slope will take $\gamma_n\leq0$ \citep{Juszkiewicz1993}.
Hence smoothing decreases values of $S_n$'s.
Thus assuming that PT results of these authors would also hold for $f(R)$ and all modified gravity effects would be encoded in the modified slope of the variance,
we can expect that the hierarchical amplitudes will be sensitive to the enhanced matter clustering exhibited by the $f(R)$ models.

\subsubsection{The estimation of moments}
\label{sub:mom_estim}

The study presented in this paper will concern the smoothed DM density fields. Having this in mind, we have decided
to design and use a special, yet simple and fast algorithm for computing the moments of the $\der$ field.  It can be summarised in
a following few steps:
\begin{enumerate} 
 \item Obtain the initial density field $\de$ on a uniform grid from a simulation snapshot using the DTFE method. This sets our limiting spatial resolution to the size of
 the grid cell. We interpolate the DTFE matter density field onto regular $N_{\rm g}=1024^3$ cubical cells grid.
 \item Perform a forward FFT ({\it Fast Fourier Transform}) of the field. Multiply the $\de(\mathbf{k})$ field values with the Fourier top-hat window for 
  a chosen value of $R_{\rm TH}$.
 \item Perform an backward FFT to obtain the smoothed real density field $\der$.
 \item Compute central moments of the distribution function using
 \ba
 \av{\der^n} = {1\over N_{\rm g}}\sum_{i}^{N_{\rm g}}\left(\der^i - \av{\der}\right)^n\,.
 \ea
 \item Finally we use the equations (\ref{eqn:cumulants}) to obtain the cumulants of the input field smoothed at scale $R_{\rm TH}$.
\end{enumerate}

By applying this algorithm to our simulation data we get the first nine cumulants of the density field for a range of
smoothing scales. We have to note that while we always use the initial number of field components $N_{\rm g}$ for each smoothing scale $R_{\rm TH}$, it is evident
that with increasing scale more and more cells will become correlated. Thus we are limited by the finite volume effects at large scales,
which becomes severe for scales $R_{\rm TH}\simgt 0.1L_{\rm box}$ \citep{Colombi1994}. At the same time we are also limited by the Nyquist sampling limit
or the initial grid spacing at small scales. Hence for the purpose of our analysis we will only consider scales that satisfy $2/\sqrt[3]{N_{\rm g}}<R_{\rm TH}/L_{\rm box}<0.1$,
where $N_{\rm g}$ is the number of the grid cells (set to be the same as the number of DM particles) and the $L_{\rm box}$ is the co-moving width of the simulation box.

The algorithm described above provides a very fast and parallelised method for obtaining higher order cumulants from a given initial density field.
We have thoroughly tested the code implementing our algorithm by comparing results with the usual spherical counts-in-cells methods \citep[\eg][]{cic_nbody1,cic_nbody2}.
For the modified gravity simulation data presented in \citep{Hellwing_npoint} we have found perfect agreement with the results both for GR and for modified gravity.

\subsubsection{Sampling errors}
\label{sub:sampling_errors}

\begin{figure}
  \includegraphics[angle=-90,width=80mm]{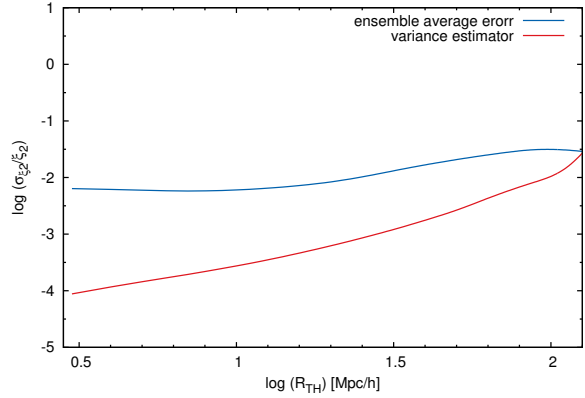}
   \caption{Comparison of two relative error estimators ($\sigma_{\overline{\xi}_2}/\overline{\xi}_2$) for the $\overline{\xi}_2$ cumulant for the GR ensemble for 1500$\hmpc$ box.
   The red line depicts the square root of variance estimator from eq. (\ref{eqn:variance_variance}), while the blue line marks the dispersion coming from ensemble average 
   of the six different realisations.
   }
 \label{fig:error_compare}
\end{figure}

In this work we focus on a direct comparison of the nonlinear clustering amplitudes between the GR (LCDM) and $f(R)$ cosmologies. In order to make the comparison 
we need to quantity the variance or the sampling errors of our measurements. The variance of a count-in-cells estimator of a cumulant of an $n$-th order, in general depends on values
of cumulants of $n+2$ and $n+1$ order \citep{1977KendallStuart}. For example the variance of the second cumulant (the field variance itself) is estimated by
\ba
\label{eqn:variance_variance}
\textrm{Var}\left[\ad{2}_{\rm c}\right]=N_{\rm g}^{-1}\left(\ad{4}_{\rm c} - 2\ad{3}_{\rm c} + \ad{2}_{\rm c} - \ad{2}_{\rm c}^2\right)\,.
\ea
In practice however the above estimator is rather cumbersome to use. This is because the high-order cumulants are more
severely affected by the finite volume effects \citep[eg.][]{Hellwing_npoint}, this effect will render eqn. (\ref{eqn:variance_variance})
unusable for cumulants of order $n\geq5$ for scales larger than $\sim20\hmpc$. For that reason we decided to use the variance of the measured cumulants
coming from ensemble averaging as our main error estimator. This is a reasonable approach,
as the errors coming from averaging between different realisations of the initial conditions are more conservative than the estimator of eqn. (\ref{eqn:variance_variance})
\citep{cic_ana, Hellwing_npoint}. To validate that we plot the Fig. \ref{fig:error_compare}, where we compare two standard deviation relative error ($\sigma_{\xi_2}/\xi_2$) estimators for $\xa{2}$.
The red line marks the relative error estimated using the eqn. (\ref{eqn:variance_variance}), while with the blue line we draw the appropriate error estimated
from ensemble averaging. The plot clearly demonstrates that the ensemble average error is conservative for all probed smoothing scales and that the both
estimators converge at large scale as expected. 

\subsubsection{Transients}
\label{sub:transients}

We would like also to discuss briefly another possible source of error in form of artificially induced bias coming from the procedure used to generate
the initial conditions for our simulations. As mentioned before we use the Zel'dovich approximation to obtain the displacement field that is used to compute
particles' peculiar velocities and displace particles from their initial Eulerian coordinates. Because the Zel'dovich procedure does not conserve momentum the density
distribution function of a field generated using this technique posses a non-vanishing artificial skewness, kurtosis and higher order cumulants. This unwanted
and unphysical deviations from the true dynamics are called {\it transients} and have been studied in detail in the literature \citep[\eg][]{transients3,transients1,transients2}.
To eliminate the effect of transients from initial conditions a system must be allowed to evolve in a pure dynamical way for a sufficiently long time. The effects of
transients for general class of models with scalar field induced fifth force was studied by \cite{Hellwing_npoint}. Their study implies that transients
effects can be of order of a few percent ($\sim 5-10\%$) for the skewness at scales where the unscreened fifth-force is allowed to act ($R_{\rm TH}\leq1\hmpc$ in their models). 

Because of the above, 
the initial redshift of a cosmological simulation is an important factor in determining the statistical reliability of the cosmological numerical experiment. 
In general, for the purpose of comparison of density fields and cumulants in different models we need to be less concerned about the net amplitude of the transients 
as they will have the same magnitude in all models. This is because in the $f(R)$ class of models we consider, the scalar field and the fifth-force have negligible
effects for the matter fields dynamics until redshifts of a few, for $z\simgt 4$ \citep[eg.][]{olh2008,LiHell2013} the growth and expansions histories are closely 
matched between GR and $f(R)$.
Therefore before the fifth force will start to change the dynamic of the density field evolution the transients will be largely erased thanks to moderately
high starting redshift of our simulations.

\section{Results}
\label{sec:results}

In this section we present analysis and discussion of main results of our study. First we focus on $z=0$ density field and its cumulants hierarchy. 
Then our analysis is followed by a detailed study of the redshift evolution of the $f(R)$ gravity effects in the clustering of the matter.

\subsection{The variance and the $\sigma_8$}
\label{subsect:variance}
\begin{figure}
  \includegraphics[angle=-90,width=80mm]{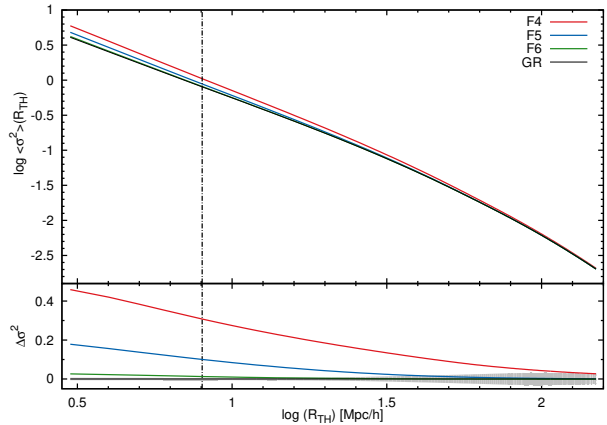}
   \caption{The average variance $\av{\sigma^2(R_{\rm TH})}$ of the density field for GR and three flavours of $f(R)$ gravity for 1500$\hmpc$ ensembles. Vertical dot-dashed
   line marks the smoothing scale $R_{\rm TH}=8\hmpc$. The shaded region represents $1 \sigma$ scatter over ensemble.
   }
 \label{fig:variance_r}
\end{figure}

First we look at the variance of the density field in our models. The two-point statistic for $f(R)$ gravity was studied both using
numerical simulations \citep{olh2008,zlk2011,lb2011,LiHell2013} and perturbation theory \citep{lb2007,fr_pertb1,fr_pertb2}. The results in the literature
mostly focus on the power spectrum of the density fluctuations $P(k)\equiv\av{\de_k^2}$. The variance of a field is related to its Fourier power spectrum by
\be
\label{eqn:sigma-pk}
\sigma^2(R_{\rm TH}) = \int{\dd k\over2\pi^2} k^2 P(k)W_{\rm TH}^{2}(kR_{\rm TH})\,.
\ee
Here $W_{\rm TH}$ is the Fourier top-hat window described by eqn. (\ref{eqn:top-hat-k}) and $R_{\rm TH}$ is the comoving smoothing scale in $\hmpc$. In cosmology the variance
of the density field plays a special role via the $\sigma_8$ parameter. The $\sigma_8$ is the square root of the density field variance smoothed with $8\hmpc$
top-hat. The linear theory prediction for the $\sigma_8$ is employed as a normalisation parameter for the power spectrum and is extensively used for generation of
initial conditions for cosmological numerical simulations. The scale of $8\hmpc$ is chosen as, in principle, for most viable cosmological models this scale
separates nonlinear density perturbation regime ($\de\gg1)$ from the linear one ($\de<1)$. In practice however these two regimes are combined by mildly nonlinear regime where $\de\sim O(1)$.
Due to existence of this intermediate regime, some mode coupling occurs and the value of the density variance at $8\hmpc$ at late times is affected by mildly nonlinear evolution.
Hence the value of $\sigma_8$ measured in cosmological N-body simulations as well as in astronomical observations is higher than the linear theory prediction \citep[for an excellent discussion see][]{RomanSimga8}.
We expect that the impact of weakly nonlinear dynamics on the variance and the value of the $\sigma_8$ parameter in particular, will be pronounced in the $f(R)$ gravity models. 
This is because, as many authors have shown, the amplitude of the density power spectrum in $f(R)$ theories is increased compared to GR for wave numbers $k\simgt 0.1 h/$Mpc.

\begin{table}
\begin{tabular}{ccccc}
\hline\hline
smoothing scale & \multicolumn{4}{c}{$\av{\sigma^2(R_{\rm TH})}^{0.5}$}\\
$R_{\rm TH}$ [$\hmpc$] & GR & F4 & F5 & F6\\
\hline
3 & $2.02$ & $2.44 (21\%)$ & $2.19 (9\%)$ & $2.05(1\%)$ \\
\hline
8 & $0.9$ & $1.03 (14\%)$ & $0.94 (5\%)$ & $0.9 (0.7\%)$ \\
\hline
20 & 0.42 & $0.45 (7\%)$ & $0.43 (2\%)$ &  0.42\\
50 & 0.173 & $0.181 (5\%)$ & $0.175 (1\%)$ & 0.173\\
100 & 0.078 & $0.08 (3\%)$ & 0.078 &  0.078\\
\hline\hline
\end{tabular}
\caption{The value of $\av{\sigma(R_{\rm TH})}^{0.5}$ for a chosen smoothing scales $R_{\rm TH}$. We do not show the values of $1\sigma$ errors from averaging,
as for small scales they are $<1\%$ and reach only $\sim3\%$ for $R_{\rm TH}=100\hmpc$. The percentage values given in parentheses are relative deviations
from the GR case, as defined in eqn. (\ref{eqn:realtive_var}).}
\label{table:variance}
\end{table}
To check how big is the effect of the scalaron for the real space density variance we present Fig.~\ref{fig:variance_r}. The top panel shows the variance $\av{\sigma^2}$
of a field smoothed over a range of scales and averaged over ensemble of six realisations of the $1500\hmpc$ simulations. The bottom panel illustrates the relative deviation
of the $f(R)$ gravity models from the values of the fiducial GR case. We define this relative deviation as
\be
\label{eqn:realtive_var}
\Delta\sigma^2 \equiv {\sigma^2_{f(R)}\over\sigma^2_{GR}}-1\,.
\ee
The lines representing different models are: black for GR, \tcr{red} for \tcr{F4}, \tcb{blue} for \tcb{F5} and \tcg{green} for \tcg{F6}. We will use this colour scheme 
throughout the paper to present our results. The black vertical dotted-dashed line marks the $R_{\rm TH}=8\hmpc$ scale, whilst the shaded region represents the $1\sigma$ scatter
around GR  ensemble mean (invisible on the top panel due to smallness of the errors). Looking at both panels we clearly see that the variance
is enhanced in $f(R)$ for a range of smoothing scales. As expected the F6 model shows weakest deviations, while the F4 exhibits strong enhancement of the clustering amplitude.
For the latter we can observe that even at scales $R_{\rm TH}\sim 100\hmpc$ the value of $\Delta\sigma^2$ is of the order of $\sim 0.05$. To allow for a better comparison between models
we show the Table \ref{table:variance}. There we examine the values of averaged standard deviation for a few chosen smoothing scales. At $R_{\rm TH}=3\hmpc$, the resolution scale of 
our $1500\hmpc$ simulations, the modified gravity effects are large for both F4 \& F5. The F6 model at this scale shows only $1\%$ enhancement of clustering amplitude.
For the F5 model values of $\av{\sigma(R_{\rm TH})}$ quickly converge to GR for scales $R\geq20\hmpc$, however the F4 model variance bears significant signal even at $50\hmpc$ and $100\hmpc$ scales.
This is emphasised by the fact that the value of nonlinear $\sigma_8$ for this model is in $14\%$ excess from GR as $\sigma_8^{GR}=0.9$ and $\sigma_8^{F4}=1.03$. 
This result could in principle be measurable, as  observational data provides estimate of the nonlinear $\sigma_8$ parameter. However for the most of the data available 
for $\sigma_8$, to be properly interpreted within $f(R)$ framework, would require some assumed model of the galaxy biasing in $f(R)$ gravity. We will address this 
issue in a forthcoming paper \citep{biasFR}. Here we can comment that some estimates of the $\sigma_8$ parameter based on peculiar velocities 
(hence largely independent on galaxy biasing) favour high 
value of this observable. For example \cite{Om_sigma8_pv} using pairwise velocities method estimate it to be $\sigma_8=1.13^{+0.22}_{-0.23}$, while \cite{s8_bulk_flow} 
by assessing the bulk flow in local universe got $\sigma_8>1.11$ at $95\%$ CL. Both velocity-based estimates are in slight tension with galaxy clustering measurements 
that usually yield lower value of the normalisation parameter $\sigma_8^{gal}=0.92\pm0.06$ \citep{galaxy_s8_1,galaxy_s8_2,galaxy_s8_3}. However to allow 
for a fair comparison with observations both methods needs to be corrected for the $f(R)$ framework. The velocity-based methods must account 
for additional accelerations induced by the fifth-force in unscreened parts of the Universe, while the galaxy-clustering method need to be corrected for 
realistic galaxy formation and biasing in $f(R)$. Both issues are subject of our work in progress and will be presented in a forthcoming paper.

\subsection{N-point functions and hierarchical amplitudes}
\label{subsect:npoint}

We move to higher order correlation functions (cumulants) hierarchical amplitudes that constitute the main subject of our study.

\subsubsection{General properties}
\label{subsect:general}
\begin{figure}
  \includegraphics[angle=-90,width=90mm]{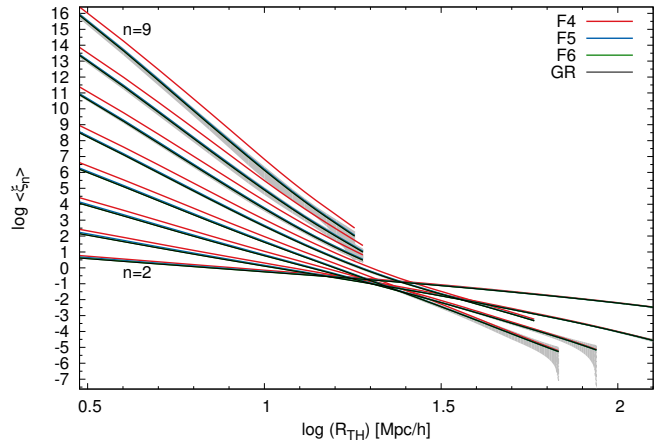}
   \caption{The averaged $n$-point correlation functions $\av{\overline{\xi}_n}$ for a range of smoothing scales. At smallest smoothing scale the lines can be clearly
distinguished by the increasing amplitude, starting from $\av{\overline{\xi}_2}$ for the lowest line, up to $\av{\overline{\xi}_9}$ for the highest amplitude. Shaded regions mark $1\sigma$ errors around the GR mean. We plot functions only out to the scales which are not yet strongly affected by the noise and finite volume effects.}
   
 \label{fig:xi-ns}
\end{figure}

\begin{figure}
  \includegraphics[angle=-90,width=90mm]{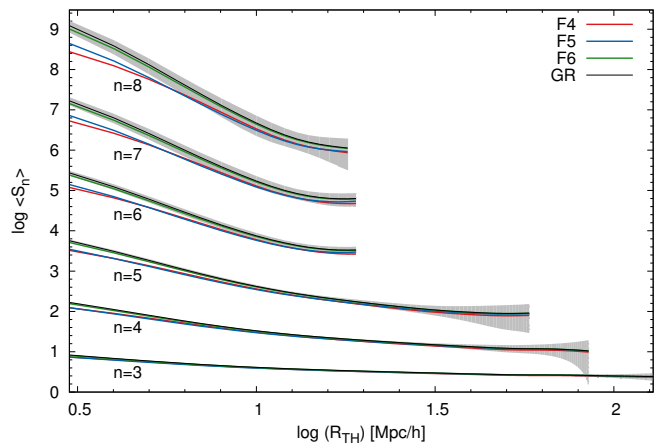}
   \caption{As in the Fig. \ref{fig:xi-ns} but this time hierarchical amplitudes $\av{S_n}$'s are displayed. Here the lines start from the
lowest order $n=3$ which marks the reduced skewness $S_3$ and are plotted for sequentially increasing order of the hierarchical amplitude up to $n=8$.}
 \label{fig:s-ns}
\end{figure}

First we take a look at the whole family of eight correlation functions from $\overline{\xi}_2$ to $\overline{\xi}_9$ and associated with them seven hierarchical amplitudes
from reduced skewness $S_3$ up to $S_8$. We plot them on the Figs \ref{fig:xi-ns} and \ref{fig:s-ns} respectively. For every GR line we also draw a shaded region that marks
the $1\sigma$ scatter around the mean value from ensemble. We note that for small and intermediate smoothing scales $3\hmpc\leq R_{\rm TH}\simlt 20\hmpc$ in the regime of
modified $f(R)$ gravity the amplitudes of volume averaged correlation functions exhibit excess when compared to the fiducial GR case. This is especially clearly seen for the F4 model.
Exactly opposite effect can be seen for the hierarchical amplitudes. Here we observe that the density field in $f(R)$ models is characterised by lower values of the $S_n$ functions
compared to the GR Universe. We also can note that the relative differences between modified gravity and GR get bigger and bigger as we move to higher and higher order amplitudes.
For $S_8$ the differences at $3\hmpc$ can be as big as a factor of a few. Fig. \ref{fig:s-ns} also illustrates the important fact, namely that in the case of $f(R)$ gravity the quasi-Gaussian 
correlation hierarchy is also present just as for the standard GR model, the main difference being that the amplitudes and their scale-dependence deviate from the standard model.

This preliminary analysis implies that we can expect to see strong modified gravity signal in the hierarchical
amplitudes at small scales, and we can expect that the relative deviation from the GR case gets stronger for higher orders. Another important observation we would like to emphasise here
regards the fact that lower values of the $f(R)$ hierarchical amplitudes actually mean that their density distribution functions are departing from their GR equivalents. We will discuss the physical interpretation of this observation later on. 

\subsection{The skewness, kurtosis and $S_5$}
\label{subsect:s3s4s5}

From the observational point of view higher-order clustering amplitudes are harder to measure and are no doubt affected by larger uncertainties. The cumulants
that are most studied for the standard gravity paradigm are skewness and kurtosis. We include also $S_5$ in this set and focus our analysis on these three first measures 
of the deviation from Gaussianity. Three-point correlations have previously been studied for modified gravity models. \cite{Bernardeau_mg} and \cite{3point_FR} studied the bispectrum,
while \cite{skewness_FR} derived the formula for the modified gravity skewness of the density field in a matter dominated (Einsten-de Sitter) Universe approximation. 
The former works find the reduced bispectrum of the modified gravity to deviate only very weakly from the GR case. For example \cite{3point_FR} for the F4 model
find deviation of the reduced bispectrum amplitude for $k\simgt0.1 h/$Mpc to be only of the order of $\sim 1\%$. Anologously \cite{skewness_FR} find deviation (lower value than in the GR)
in the reduced skewness to be at best of the order of $\sim 2\%$ for a strongly coupled scalar field. Both results were obtained using perturbation theory that includes
second order terms. The validity of such approach was largely tested for the three-point statistic in the GR universe. Surprisingly many authors 
\citep[eg.][]{Juszkiewicz1993,grav_inst_pt2,grav_inst_pt3} found good agreement with N-body simulations also in the regime where weakly non-linear perturbation 
theory should fail, i.e. $\de\sim1$. However as we will see later on this approach fails for the modified gravity models. First of all the class of the modified gravity 
theories and the $f(R)$ models we study here in particular, are characterised by the higher degree of nonlinearity. This is due to stronger clustering as showed by us for 
the case of variance in \S\ref{subsect:variance}, but also the highly nonlinear character of the evolution and distribution of the chameleon field adds to 
the total degree of nonlinearity of the density field. Secondly, the perturbation
theory is limited and ill-posed to look for the modified gravity signatures in the $\delta$ field, as these signatures are strongest at small scales, which by construction are
beyond the validity of the perturbation regime.
\begin{figure}
  \includegraphics[angle=-90,width=80mm]{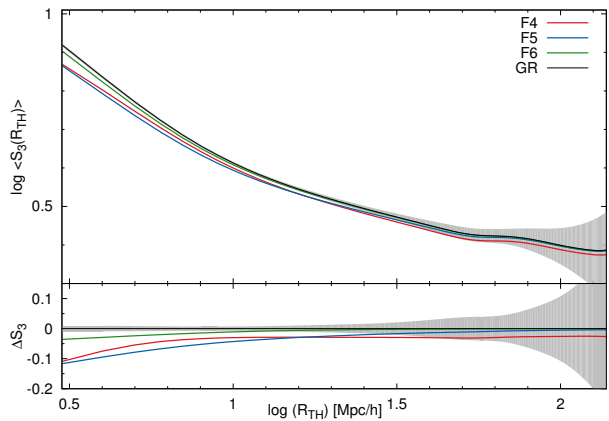}
   \caption{The averaged reduced skewness $\av{S_3}$ form 1500$\hmpc$ ensembles. The solid colour lines depict ensembles: GR (black), F4 (red), F5 (blue) and F6 (green). 
   The lower panel presents the relative difference $\Delta S_3$ from the GR case. The shaded area illustrate $1\sigma$ scatter around the ensemble mean for the GR simulations.}
 \label{fig:skewness}
\end{figure}

In Fig. \ref{fig:skewness} we plot the real space skewness obtained from the $1500\hmpc$ ensemble. The top panel shows the absolute value of the $S_3$ for the field
smoothed at a range of scales $3-150$ $\hmpc$. The bottom panel illustrates the relative deviation from the GR values $\Delta S_3$, defined in the analogous way as in 
eqn. (\ref{eqn:realtive_var}). As usual the shaded regions quantify the $1\sigma$ deviations from the GR mean. In connection with the above-mentioned results of other
authors we indeed confirm that at large scales $R_{\rm TH}\simgt 40\hmpc$ both F5 and F6 models converge to the GR case. However this is not the case for the F4 model.
The skewness in this model bears the signal of modified dynamics at level of $\sim 5\%$ from $10$ up to $100\hmpc$. We also find that there is regime of strong
deviation of $f(R)$ gravity clustering from the Einstein's theory case. It appears for the scales $\leq10\hmpc$. We find the strongest signal at the resolution 
limit $R_{\rm TH}=3\hmpc$ of our $1500\hmpc$ box simulations. Here $\Delta S_3=-12\%$ both for F4 and F5 models and is still of order of $-4\%$ for the F6 case. We would like also to make a side
remark on this occasion. In our $S_3$ data we have found the BAO ({\it Baryon Acoustic Oscillations}) signal for all models at scales predicted by \cite{SkewBAO,SkewBAO2}. The wiggle can be clearly seen on the fig. \ref{fig:skewness} for scales $1.5<$log$(R_{\rm TH})<2$.
\begin{figure}
  \includegraphics[angle=-90,width=80mm]{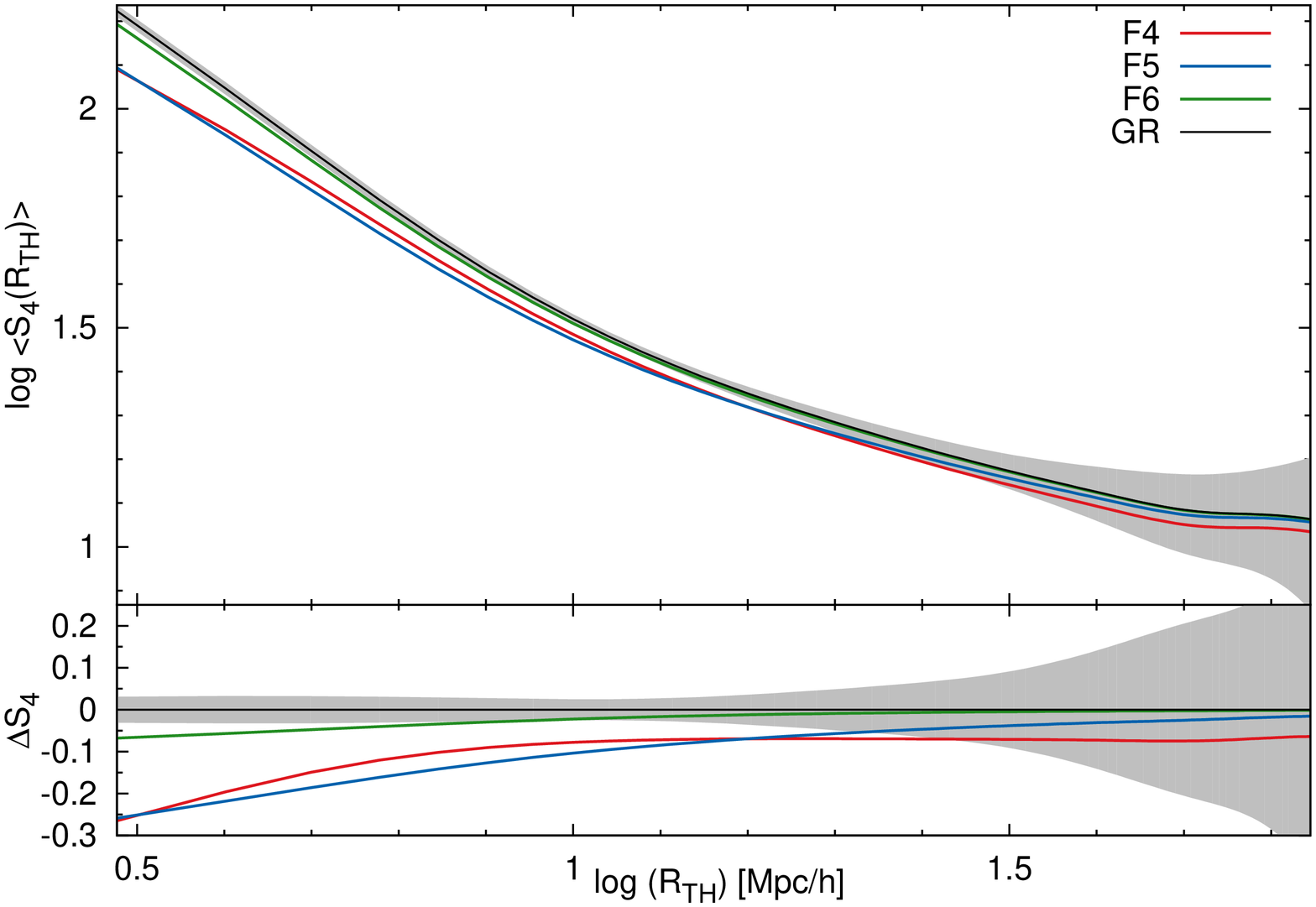}
   \caption{The averaged kurtotsis $\av{S_4}$ from 1500$\hmpc$ ensemble. The defintions of lines and panels as in the Fig. \ref{fig:skewness}.}
 \label{fig:kurtosis}
\end{figure}
\begin{figure}
  \includegraphics[angle=-90,width=80mm]{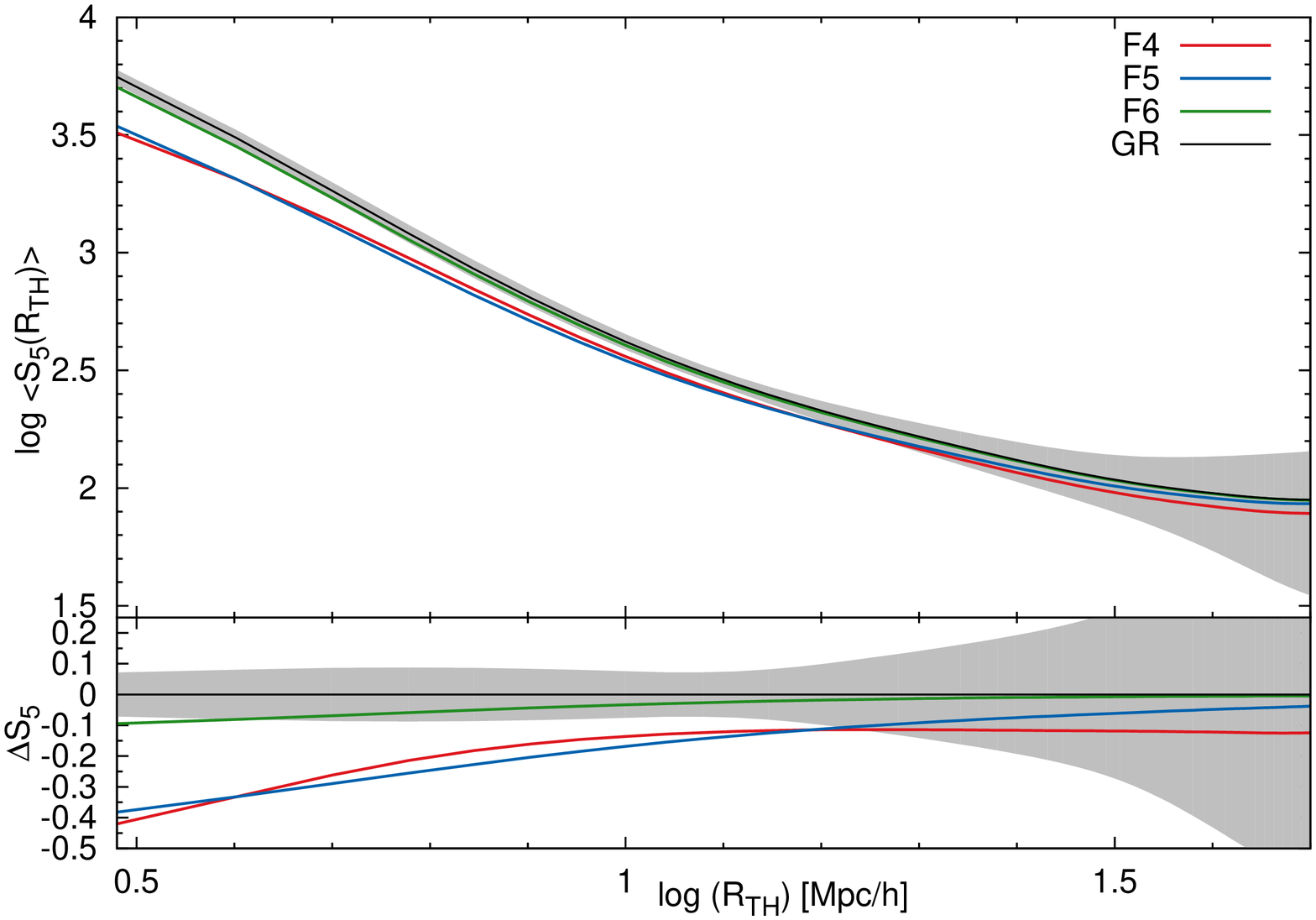}
   \caption{The averaged $\av{S_5}$ from 1500$\hmpc$ ensemble. The defintions of lines and panels as in the Fig. \ref{fig:skewness}.}
 \label{fig:s5}
\end{figure}
\begin{table*}
\begin{tabular}{cccc}
\hline\hline
smoothing scale & S3 & S4 & S5 \\
$R_{\rm TH}$ [$\hmpc$] & \multicolumn{3}{c}{GR,\tcr{F4},\tcb{F5},\tcg{F6}}\\
3 & $8.31\pm0.08$, \tcr{7.4}, \tcb{7.34}, \tcg{8.01} & $167.2\pm5.2$, \tcr{123}, \tcb{124}, \tcg{156} & $5661\pm403$, \tcr{3269}, \tcb{3491}, \tcg{5123} \\
8 & $4.53\pm0.03$, \tcr{4.38}, \tcb{4.3}, \tcg{4.47} & $42.5\pm1.2$, \tcr{38.7}, \tcb{37.1}, \tcg{41.2} & $643\pm54$, \tcr{540}, \tcb{512}, \tcg{615}\\
20& $3.3\pm0.04$, \tcr{3.21}, \tcb{3.22}, \tcg{3.29} & $19.2\pm0.9$, \tcr{17.9}, \tcb{18.1}, \tcg{19.1} & $165\pm23$, \tcr{146}, \tcb{150}, \tcg{163}\\
50& $2.7\pm0.1$, \tcr{2.61}, \tcb{2.67}, \tcg{2.69} & $12.2\pm2.5$, \tcr{11.2}, \tcb{11.9}, \tcg{12.1} & $89\pm54$, \tcr{78}, \tcb{86}, \tcg{89}\\
100&$2.51\pm0.27$, \tcr{2.44}, \tcb{2.49}, \tcg{2.51} & $9.5\pm17.2$, \tcr{8.7}, \tcb{9.3}, \tcg{9.4} & -, \tcr{-}, \tcb{-}, \tcg{-}\\
\hline\hline
\end{tabular}
\caption{The values of averaged hierarchical amplitudes for $n=3,4$ and 5 presented here for a few chosen smoothing radii. Each column contains four
comma-separated numbers. The first one (black) gives the GR mean plus error, the second (red) is for the F4 model, the third (blue) corresponds to F5 and
finally the fourth (green) represents F6.}
\label{table:s3s4s5}
\end{table*}

The mark of modified gravity is further enhanced in the case of the kurtosis and the fifth-order amplitude $S_5$. We plot corresponding data on figures 
\ref{fig:kurtosis} and \ref{fig:s5}. For our limiting radius of $3\hmpc$ the relative deviation from the fiducial GR case reaches $\Delta S_4=-26\%$ and 
$\Delta S_5=-42\%$ respectively. Furthermore the statistical significance of the measured deviations is very big. For the $R_{\rm TH}=3\hmpc$ the F4 mean values
are 11, 8.5 and 6.9 $\sigma$ away from the GR mean for $S_3$, $S_4$ and $S_5$ respectively. This means that in statistical sense the density field at those scales is characterised by different shape
density distribution functions for each of our models. For a better comparison we have collected the values of measured $S_3,S_4$ and $S_5$ for a few chosen
smoothing scales in the Table \ref{table:s3s4s5}.

\begin{figure}
  \includegraphics[angle=-90,width=80mm]{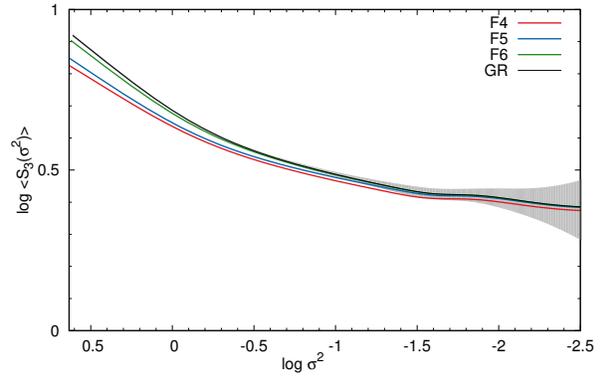}
   \caption{As the figure \ref{fig:skewness}, but this time the averaged skewness is plotted against averaged field variance $\av{\sigma^2}$}
 \label{fig:skewness-var}
\end{figure}
Closer inspection of the data plotted in the Figs \ref{fig:skewness}, \ref{fig:kurtosis} and  \ref{fig:s5} reveals an interesting feature. It appears that 
for the scales below $\sim10\hmpc$ the relative deviation is {\it stronger} for the F5 model than for F4 model. At our resolution limit both signals 
seem to converge to a similar value. However for the mentioned scales the hierarchical amplitudes of the F5 density field are smaller than for any other 
considered models. To understand better this behaviour we show the Fig. \ref{fig:skewness-var}, where the skewness is plotted against variance $\sigma^2$ 
of the field rather then smoothing scale. The plot shows that for all values
of the variance the lowest skewness belongs to the F4 model, as we would initially expect. At small scales, for the same smoothing radius it is the F4 model
that has the largest variance. Hence the lines from Fig. \ref{fig:skewness} get shifted accordingly. For the general class of the fifth-force cosmology \cite{Hellwing_npoint} found
that the stronger the fifth-force or larger screening length the stronger the deviation in the $S_n$'s functions. This is apparently not the case for our F5 and F4 models.
We can propose the following explanation of this phenomena. In generic flavours of modified gravity models the fifth force is allowed to act freely on all the scales
of interest for a particular model. Thus once the force arises due to non-minimal coupling of the scalar field to matter it changes the dynamics of the matter field. In realistic
family of such models the fifth-force is usually suppressed for large scales and can only act on small and intermediate scales. Therefore an unavoidable quality 
of a model without environment dependent screening is that the effects of modified gravity are strongest at small scales. Now if we consider our $f(R)$ models, we need to take into account 
the chameleon mechanism that is
screening out the fifth-force in dense parts of the field. 

Now we can naturally explain the unexpected behaviour of the $S_n$'s functions for the F4 and F5 models. The F4 model
is the one that experiences the strongest clustering. As mentioned before it has the non-linear $\sigma_8^{F4}=1.03$, which should be compared with $\sim9\%$ lower 
value of the F5 model $\sigma_8^{F5}=0.94$. Due to stronger clustering and more efficient matter accretion cluster mass haloes get more massive \citep{sloh2009}. In general we can expect that,
on average, the small-scale matter aggregations like clusters and filaments will be denser in the F4 model when compared to F5. 
In addition, in the F4 model the chameleon screening is much less effective when compared to the F5 universe. Hence while both F4 and F5
models experience fifth-force enhanced dynamics in low density environments like cosmic voids and walls, the fifth force in the F5 model
is partially screened out in dense cluster and filaments.
This naturally leads to stronger deviation
in the hierarchical amplitudes at small scales. To conclude, what we observe in the behaviour of the $S_3, S_4$ and $S_5$ values shown in Figs \ref{fig:skewness}-\ref{fig:s5}
is the chameleon mechanism caught in the act during large-scale structure formation in $f(R)$ gravity models.

\subsection{The redshift evolution}
\label{subsect:z_evol}
\begin{figure}
  \includegraphics[angle=-90,width=80mm]{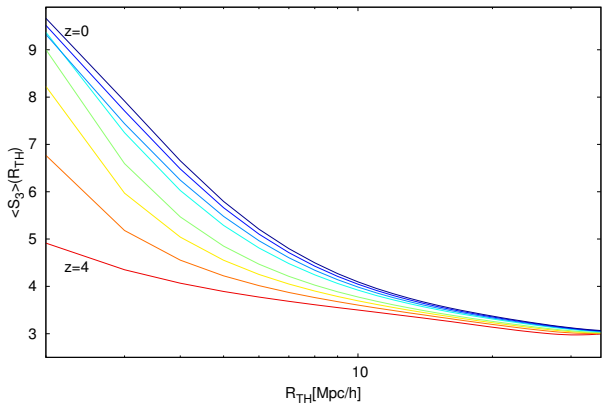}
   \caption{Gravitational instability at work. The redshift evolution of the skewness for the GR model in 1000$\hmpc$ box. The lines from bottom (red)
    to top (yellow) mark snapshots taken at consecutive redshifts: $4,2.33,1.5,1,0.43,0.25,0.1$ and 0.}
 \label{fig:red_evo_s3}
\end{figure}
 
So far we have focused on the $z=0$ results of our simulations. From physical, but also observational point of view, it is important to study the time evolution
of the high-order correlation hierarchy. The hierarchical structure formation paradigm that is a part of the cold dark matter model assumes that structures in the Universe
arise from primordial tiny Gaussian-distributed density fluctuations by means of the gravitational instability mechanism. In this picture the high redshift density
distribution function is closer to a Gaussian, and as the system is evolving in time the gravitational dynamics drives the density field far away from the
initial Gaussian distribution. The prediction is that the skewness and higher order amplitudes grow with time at scales where the non-linear and mildly non-linear 11
evolution occurs \citep[see \eg][]{skewness_barions}. Since as we know in the $f(R)$ cosmology the density field experiences a modified (enhanced) density perturbation growth history,
we also expect that the pattern of growth in time of the hierarchical amplitudes will be modified w.r.t. the GR case.

\begin{figure*}
  \includegraphics[angle=-90,width=170mm]{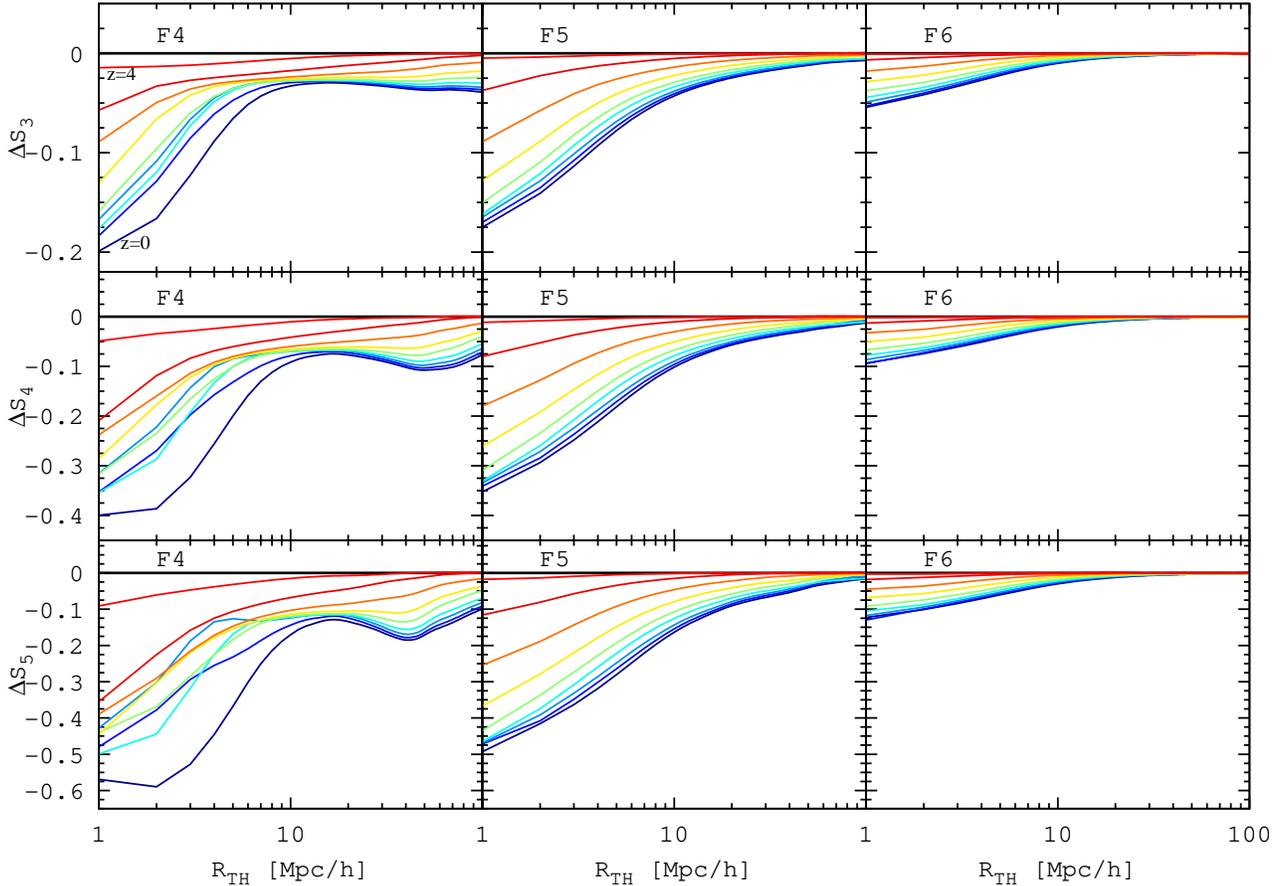}
   \caption{Time evolution of the skewness (top-row panels), the kurtosis (middle-row panels) and $S_5$ (bottom-most panels) deficiency
with respect to the fiducial GR model. The left-most columns show the F4 model, the centre columns marks the F5, while the right-most panels corresponds to the F6 flavour. The data is from 1000$\hmpc$ ensemble.}
 \label{fig:red_evo_s345}
\end{figure*}
We begin by looking more closely at the time evolution of the skewness for the GR model. In Fig. \ref{fig:red_evo_s3} we plot the averaged smoothed skewness for 
the GR 1000$\hmpc$ box simulation at 9 different time steps. The lines are coloured according to redshift with the reddest marking the highest redshift $z=4$ 
and the bluest pointing to $z=0$. In the  considered redshift range we observe that the values of the skewness converges for $R_{\rm TH}\geq35\hmpc$. However at smaller scales
a much higher positive skewness develops with time. This effect is of course driven by non-linear gravitational evolution, namely collapse of DM haloes and emptying
of cosmic voids. Here at $R_{\rm TH}=2\hmpc$ the skewness doubles its value from $S_3=4.9$ to $S_3=9.5$ between $z=4$ and $z=0$. We also denote that the fastest growth of
the skewness appears at earleir times, for $z>1.5$. Below this redshift the skewness grows much slower.

Once we have established what the time evolution of the skewness looks like for the LCDM model we are now ready to quantify the redshift evolution of the relative deviations 
of the $f(R)$ skewness, kurtotsis and $S_5$ from the standard gravity model predictions. This is illustrated by Fig. \ref{fig:red_evo_s345}, where we plot the
time evolution of $\Delta S_3, \Delta S_4$ and $\Delta S_5$ for the F4, F5 and F6 models. The three columns on this figure correspond to our three flavours of the $f(R)$,
from F4 (the most-left columns) to F6 (the most-right column), while the different rows present consecutive hierarchical amplitude deviations from $\Delta S_3$ in the top row to
$\Delta S_5$ presented at the bottom row. We start our analysis by looking at the F5 and F6 models, putting aside the strongest F4 case for the moment, as it experiences the most complicated 
time evolution. Again the lines are coloured according to the corresponding redshift, starting from the $z=4$ for the reddest line down to $z=0$ marked by the bluest colour.
First we denote that the F6 model data looks like a weaker and retarded in time version of the F5 model . General trends are the same for both models. The deviations 
from the GR density field undergo the fastest evolution for $1\simlt z\leq 4$. At redshift $\sim 0.6$ most of the difference between the F5, F6 and the GR Universes
is already in place and for the remaining expansion history of the Universe the deviations of both models' hierarchical amplitudes grows only weakly. These tendencies
are somewhat similar also for the F4. Here though we observe a development of a very interesting pattern of deviations in time and scale. First of all we observe that the
the $\Delta S_n$s functions are no longer monotonic with time and scale. While the deviations are still strongest at small, non-linear, scales $R_{\rm TH}\leq 10\hmpc$,
we see that for the scales $10\simlt R_{\rm TH}/\hmpc\simlt 50-60$ the actual values of the $\Delta S_n$s grow slowly with the scale. This behaviour occurs only for $z\leq1$
and is not present at higher redshifts where the patterns are similar to the F5 and F6 models. We can also discern a dip developing with time at scales $\sim 45-50\hmpc$,
with the scale of the dip being smaller for higher order amplitude. The dip is weakly visible for the $\Delta S_3$, albeit it gets much stronger and clearer
for the kurtosis and especially the $S_5$ case. The scale of this feature coincides strongly with the scale of the first dip from the BAO wiggle observed in hierarchical amplitudes of the fiducial GR model by \citet[see in the figure ~1]{SkewBAO2}. Thus we speculate that due to the extended non-linear evolution
observed in the F4 model, the BAO signal naturally present at this scales in hierarchical amplitudes \citep{SkewBAO} gets enhanced in our strongest $f(R)$ model.
This behaviour can have potentially observable consequences as the BAO signal will be measured with a percent-level accuracy in forthcoming galaxy redshift surveys.
Alas the detailed analysis of this phenomena lies beyond the scope of this paper and we leave it for the future work. 

Finally we note the complicated pattern of
the deviations in time evolution seen at small scales ($R_{\rm TH}<10\hmpc$). In contrast with the F5 and F6 models at those scales the relative deviations keep growing
steadily also at late times $z\leq 0.6$. Moreover we observe that for a few redshifts and scales the monotonic time dependence of the growth of the relative deviations is broken.
This is clearly visible for $\Delta S_4$ and $\Delta S_5$, as we can find snapshots where the values at $R\sim5\hmpc$ are actually higher for early times (eg. $z=0.6$) 
rather than for later (eg. $z=0.25$). We assign this complicated and unexpected, at first sight, behaviour of the density field distribution functions with the low efficiency of the chameleon mechanism in the F4 model. Due to this the fifth-force in this model is largely unscreend for all scales of interest. This induce much higher order of non-linearity of the density field in the F4 model (as already indicated by much higher variance) and makes the transfer of the power from small to large scales more efficient. As this non-linear 
{\it tenue} is strongly enhanced for the F4 model thus is not observed for the F5 and F6 $f(R)$ models, where the chameleon and scalaron mechanism are subject to a more graceful evolution.

\subsection{Density distribution functions}
\label{subsec:pdf}
\begin{figure*}
  \includegraphics[angle=-90,width=170mm]{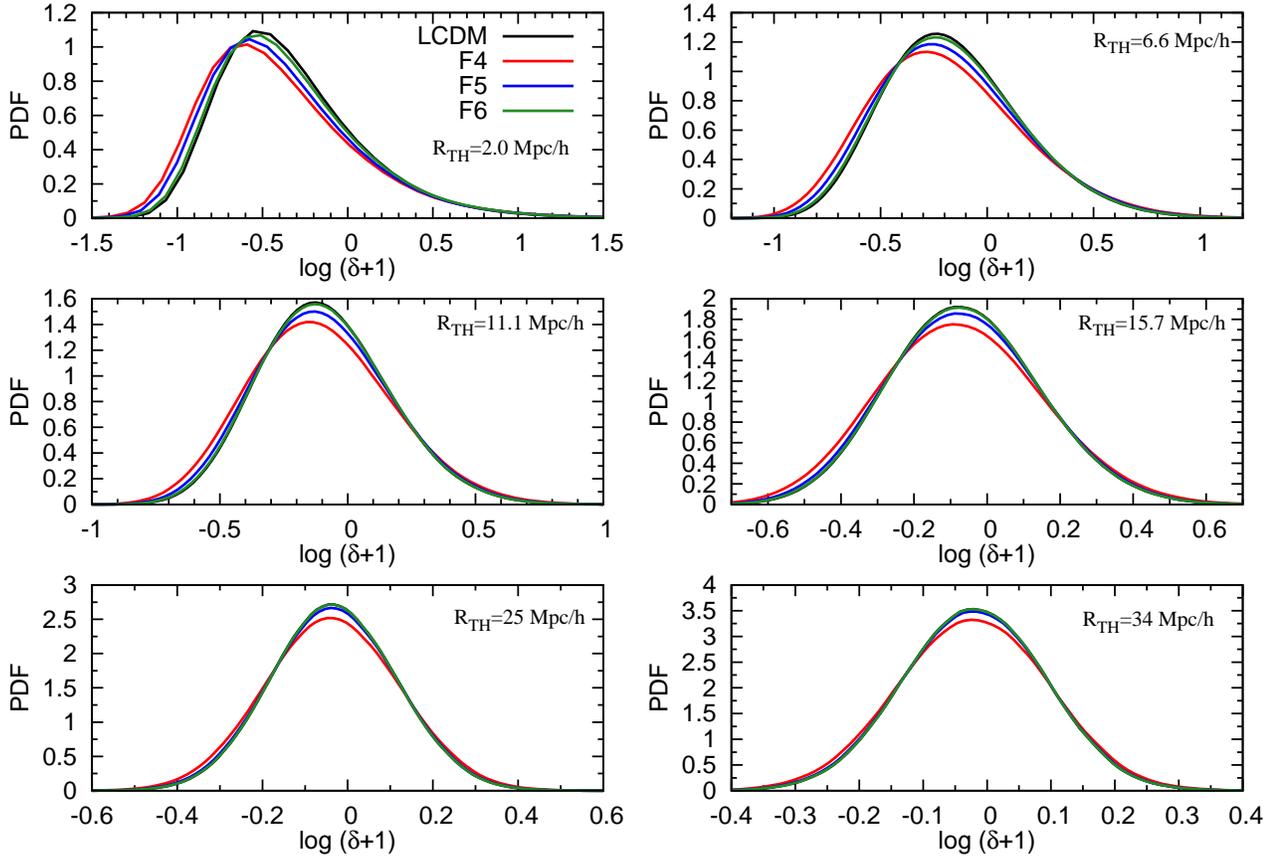}
   \caption{The probability distribution functions (PDFs) for the density field $\de+1$ computed using various top-hat
   windows. For each panel we plot PDFs for four our models marked by solid colour lines as: GR (black), F4 (red), F5 (blue) and F6 (green).
   From top to down and left to right the panels illustrate the distribution functions of a field smoothed at larger and larger scales,
   starting from $R_{\rm TH}=2\hmpc$ for the top-left panel and ending at $R_{\rm TH}=34\hmpc$ for the bottom-right one.
   See the text for more details. 
   }
 \label{fig:pdfs}
\end{figure*}

In the previous paragraphs we have assessed the patterns of the $f(R)$ gravity visible in the hierarchical amplitudes and their time evolution
as the relative deviations from the values predicted for the standard gravity limit. As discussed before for a Gaussian random field
all cumulants of the density field probability distribution function (PDF) except for the variance vanish. The non-zero higher order
cumulants of the PDF measure the shape deviations of the distribution function from a Gaussian. Thus in this section we look directly 
at the PDFs computed for all our models for fields smoothed with six different top-hat radii. For completeness we recall the definition
\be
\label{eqn:pdf_def}
\textrm{PDF}(\eta) = {P(\eta)\over \textrm{d}\eta}\cong {P(\eta)\over\Delta\eta}\,\,,
\ee
where $\eta\equiv\de+1$ and the probability $P(\eta)$ is measured in practice as a frequency probability using
\be
\label{eqn:num_prob}
P(\eta)= {1\over N_{\rm g}}\sum_{i=0}^{N_{\rm g}}(\eta_i: \eta<\eta_i\leq\eta +\Delta\eta)\,.
\ee
Here the $N_{\rm g}$ is the number of the grid points sampling the smoothed density field and we choose to keep the bin width 
$\Delta\eta$ constant in the logarithmic space.

In Fig. \ref{fig:pdfs} we present the PDFs computed for the $1000\hmpc$ box simulations, 
where in each panel the black solid line marks the GR value,
while the red, blue and green solid lines correspond to the F4, F5 and F6 models. We start from the $R_{\rm TH}=2\hmpc$ 
for the top-left panel and increase the window size from left to right and from top to bottom, ending with $R_{\rm TH}=34\hmpc$
shown in the bottom-right panel.  The general trends abide the $f(R)$ PDFs that are shifted towards lower density
contrast values when compared with the GR. This shift is accompanied only by a very small excess in the positive tail
of the density distribution. This excess is barely visible in our plot as we choose to use linear rather than logarithmic
scale on our x-axis. Commonly adopted convention is to plot PDF in the logarithmic space. We drop the logarithmic
scaling on the PDF values since it over-represents rare events (cells with very high $\de+1$ values).  By construction our
DTFE-estimated density field is volume-weighted (recall the \S\ref{subsect:density}), and as the volume of the Universe is 
dominated by cosmic voids \citep{cosmicweb,cosmicweb2,cosmicweb3,nexus}, it is clear that the modified gravity signal 
we see in the dislocated shape of  the PDFs is coming mostly from the cosmic voids. The last statement also 
holds for the patterns we saw in the reduced skewness, kurtosis and $S_5$.  The departure of $f(R)$ distribution 
functions from the GR case is quickly decreasing with the increasing radius of the smoothing top-hat, in agreement 
with the behaviour observed earlier for the hierarchical amplitudes. At scales $R_{\rm TH}\sim25-34\hmpc$
the F5, F6 and GR PDFs become indistinguishable. The strongest F4 model still bears some signal, although it is
mostly contained in a modified PDF amplitude at the mean-field values (around $\de\sim0$) rather than the PDF shape.
The deeper voids (less dense) are characteristic mark for the modified gravity models employing the scalar fifth-force.
This was established by many authors \citep{halo_void_fr,mog_void_exset,scalar_voids_Li,rebel1}. As we do not employ any 
void-finding algorithms in our studies, we probe the
void population indirectly, only in statistical sense, by measuring the density field PDFs for various radii.  Still we
can confirm that, when one is concerned with the density field clustering statistics, it is the cosmic voids that are the most
sensitive parts of the cosmic web, provided that one is interested in the modified gravity effects. 
This is especially true for the chameleon class of $f(R)$
gravity models that we study here. This is due to the fact, that the low density in voids prevents the chameleon
mechanism from screening-out the fifth-force of the scalaron at later stages of evolution \citep{scalar_voids_Li, mog_void_exset}. Hence
the fifth-force in deep voids is allowed to quickly saturate to its maximum enhancement value of $1/3$.

The $f(R)$ gravity effects seen in the PDFs from Fig. \ref{fig:pdfs} indicate that in statistical sense we have many more
parts of the density field where the density contrast is lower than in the GR case. Since our simulations share the same
initial phases the density fields both in $f(R)$ and GR share the same number of initial peaks (haloes) and dips (voids). At
 small scales the non-linear evolution can significantly alter the number of peaks and dips due to halo and void mergers as well as
the {\it void-in-cloud} process \citep[void squashing, see][]{voidado1,voidado2,JenningsVoid}. However at larger scales beyond the cluster sizes
$R_{\rm TH}\simgt5\hmpc$ this non-linear processes are much less important and do not affect much the initial peak/dip counts \citep{BBKS}.
 Therefore starting from the smoothing radius $R_{\rm TH}=6.6\hmpc$
(the top-right panel on the figure) the density fields of the GR and $f(R)$ simulations should have roughly closely matching counts statistics
of the number of peaks and dips. If this hypothesis is true the PDF shifts observed in the Fig. \ref{fig:pdfs} would indicate that
in $f(R)$ gravity we deal with cosmic voids that are emptier than their GR cousins. Since during the evolution of the Universe the
continuity equations holds everywhere (conservation of mass), the enhanced emptying of the voids must be followed by an increased
mass of the cluster and galaxy haloes. This was already observed \citep{sloh2009,halo_void_fr}. 

We can now understand why the $f(R)$ gravity clustering hierarchy has PDFs  characterised by lower values of the reduced cumulants than the GR case.  Let's  consider the gravitational instability mechanism that governs the shape
evolution of the density distribution function. As the voids get more empty the matter flow via walls and filaments towards
dense nodes of the cosmic web - the galaxy clusters \citep{cosmicweb}. This process naturally lowers the density contrast in voids,
while raising it in the more dense clusters and filaments. The lower density $\de<0$ tail of the distribution function
is constrained from the left-hand side by a natural physical limit of $\de>-1$. The density contrast can not be lower than this, since
$\de=-1$ already indicates empty space. At the same time the right-hand tail of the PDF can grow towards arbitrarily high
$\de$ values as the matter accretion, halo mergers and violent relaxation, proceed. This asymmetry, intrinsically connected with
the gravitational instability mechanism, implies growth with time of the skewness, kurtosis and higher-order cumulants. The enhanced
clustering exhibited by the $f(R)$ models provides much empty voids and this process dominates over the mass increase of
clusters (due to the volume dominance of the voids). The overall effect shifts the PDF towards the lower density tail.

\section{Concluding Remarks}
\label{subsect:conclusions}

Having assessed all parts of the dark matter hierarchical clustering in chameleon $f(R)$ gravity we can conduct
the summary of our results. We begin by noting that both fiducial GR (LCDM) model and our
three flavours of $f(R)$ models start from the same initial conditions, currently tightly constrained by high-redshift
Universe observations like the Cosmic Background Radiation temperature and polarisation anisotropy maps 
\citep{WMAP9,Planck1}. The starting point for all our models is the primordial
post-inflationary density perturbation field that has Gaussian statistics. The primordial field is then
propagated using the Zel'Dovich approximation down to the initial redshift of $z_{\rm ini}=49$.
The transients induced by this procedure are relaxed by the real dynamics followed by the
N-body simulations. In addition, the background cosmology
as well as the expansion history is shared among all models down to $z=0$. Albeit the growth history of
the $f(R)$ models traces the GR values only for $z\simgt5$, for the remaining part of the cosmic evolution
the fifth-force starts to modify the matter dynamics. Thus all departures of the clustering statistics present
in the $f(R)$ universes must be accounted for by the late-time altered dynamics induced by the modified
gravity of those models.

The effects of the scalar force dynamics are not trivial to predict,
especially at later evolutionary stages. This is due to the presence of the chameleon mechanism and  its non-linear nature
reciprocally connected with high-density peaks (haloes). 
The highly non-linear nature associated with the chameleon screening quickly renders
all perturbation theory results  inaccurate in their description of the clustering statistics
for the class of $f(R)$ models studied here \citep[\eg][]{lz2009,LiHell2013,halo_void_fr}. Our employment of the series of high-resolution, state-of-the-art
N-body simulations of the chameleon $f(R)$ gravity allowed us to perform a robust and consistent analysis.

Using our simulations we have constructed high-resolution volume weighed DTFE density fields and have used
these to compute the high-order correlation functions and hierarchical amplitudes up to respectively 9th and 8th order.
We have also traced the time evolution of the reduced cumulants focusing on the skewness, kurtosis and $S_5$ amplitudes
and the redshift-evolution of the relative deviation of these quantities from the fiducial GR case. 
Finally we have computed the density probability distribution functions for a set of the smoothing scales. We can
summarise our findings in the following points:
\begin{enumerate}
 \item The $f(R)$ density fields are characterised by higher variance $\sigma^2$. The deviation from the GR case
 is strongest at the smallest scales reliably probed by our simulations $R\sim 2-3\hmpc$ and can reach nearly $\sim 50\%$ of enhancement for F4 model.
 The effect is weaker for F5 ($\sim20\%$ maximal enhancement) and marginal for F6. The excess of the variance
 quickly drops with the smoothing scale. However for the F4 it is still of the order of $5\%$ at 100$\hmpc$.
 \item Increased variance of the DM density induces higher $\sigma_8$ values for the modified gravity. We can report that,
 the non-linear $\sigma_8$ is higher by $14\%$ in F4, $5\%$ in F5 and $0.7\%$ in F6 models and this excess from the GR is statistically significant.
 \item All measured volume-averaged correlation functions $\overline{\xi}_n$, up to 9th order have higher values for chameleon gravity at small, non-linear scales.
 However above $R>\sim 30\hmpc$ for all our $f(R)$ models the correlation functions start to converge to the GR values within  $1\sigma$  cosmic variance scatter.
 \item We have found that the hierarchical scaling is also present in the $f(R)$ gravity. The modified dynamics induced by the fifth-force 
 changes however the values  of the hierarchical amplitudes $S_n$'s and their scale dependence w.r.t. the standard gravitational instability predictions.
 The values of the modified gravity scaling amplitudes are always {\bf lower} than in the GR case. In case of the F4 and F5 models the lower
 values of the skewness, kurtosis and $S_5$ appear for all smoothing scales probed by us, up to $100\hmpc$.
 \item We have measured an interesting behaviour of the relative deviations for $\Delta S_{3,4,5}$ at scales $R\simlt 10\hmpc$ for the F5 and F4 models.
 At those scales the F5 model shows higher deviations form the LCDM case than the F4 model. We attribute this behaviour to the non-linear  evolution of the density field and the screening effect of the chameleon, which is present in the F5 and barely active in F4. Also for all models the relative deviations grow with the increasing order of the $S_n$.
 \item The evolution of the $\Delta S_{3,4,5}$ is monotonic in time. For $z>5$ the departures from the standard model are negligible and start
 to grow quickly for later times. All models exhibit the fastest growth at moderate redshifts $1\leq z\leq 4$. At late evolutionary stages, the
 departures from the GR experience much steadier growth. This picture however becomes much more complicated for the strongest F4 model.
 Here we have observed a highly nonlinear pattern of scale and time dependence of the relative deviation parameters $\Delta S_{3,4,5}$.
 We believe that this effect is due to severely enhanced non-linearities in the density field present in the F4 universe.
 \item The probability distribution functions of the $f(R)$ density fields are significantly shifted towards the $\de\rightarrow -1$ tail.
 This can be observed for scales up to $R_{\rm TH}\sim 20\hmpc$. Hence in a statistical sense the $f(R)$ gravity produces much emptier and deeper
 cosmic voids.
 \end{enumerate}
 
 Our findings summarised above paint a picture in which the $f(R)$ gravitational instability and dynamics induce significant
 differences in the degree of the dark matter density field correlations at small scales. The most important
 message is that the hierarchical scaling is preserved in this class of models as well. Albeit we can denote that at scales relevant to the
 galaxy and halo formation, the $f(R)$ density field is characterised by enhanced clustering at all correlation orders.
 
 Before any comparison of our theoretical predictions to observations can be made, we need to briefly address three
 possible sources of confusion.
 
 {\it 1. Galaxy biasing.} If the galaxy formation process varies in efficiency with environment, then the galaxy distribution 
may foster a biased picture of the underlying mass distribution. If we assume that the smoothed galaxy density field $\de_{\rm g}$
is a local, but not necessarily linear function of the smoothed DM density field: $\de_{\rm g}(x) = F[\de(x)]$, where $F$
is an arbitrary function, then by using a Taylor expansion of $F$ for $|\de|\ll1$, it can be shown, as \citet{FG1993}
did, that local biasing preserves the form of the scaling relations (\ref{eqn:h-scalling}). As the scaling relations for
the galaxy density field are preserved, the values of the hierarchical amplitudes $S_n$ are necessarily not. If we
consider for example the reduced skewness $S_3$, the galaxy biasing will change it to \citep{FG1993}
\be
\label{eqn:s3_bias}
S_{\rm 3g} \equiv \av{\de_{\rm g}^3}/\av{\de_{\rm g}^2} = b_1^{-1}S_3 + 3b_2/b_1^2\,,
\ee
where $b_n\equiv$d$^nF/$d$\de^n$, evaluated at $\de=0$. The bias parameters can be also scale dependent
(and in realistic models they indeed are). Generally the $f(R)$ model will differ from the GR by
the values of the $b_n$ parameters. In the worst case scenario all clustering effects induced by the
chameleon $f(R)$ gravity may be camouflaged as the scale dependent relative bias differences $b_n^{f(R)}/b_n^{GR}$,
and indeed this was already shown to some extended by \citet{sloh2009}.
Alas both the GR and $f(R)$ gravity can then predict the same values of $\av{\de_{\rm g}^n}_{\rm c}$ and $S_{n\rm g}$'s. Hence
the modified gravity signal will be visible only in a difference in the scale dependence of the bias parameters to
that in the GR.
The detailed discussion of this situation is beyond the scope of this work. We will analyse it in the forthcoming paper \citep{biasFR}.

 {\it 2. Redshift space distortions.} This is yet another important difficulty which needs to be considered before
the theoretical predictions can be meaningfully compared with the data from galaxy redshift surveys. In such catalogues
radial velocities of galaxies are used instead of their true radial coordinates. As a result, peculiar motions of galaxies
distort their true spatial distribution and the $n$-point correlation functions \citep[\eg][]{FG1994}. Both \citet{grav_inst_pt2}
and \citet{grav_inst_pt4} using the Lagrangian perturbation theory and N-body simulations, studied the effects of the redshift distortions on the $\xi_2$, $\xi_3$ and $S_3$. 
They showed that albeit both $\xi_2$ and $\xi_3$ are affected by the redshift space distortion, all appreciable effects
cancel out for the reduced skewness $S_3$. Thus, the comparison between theoretical predictions and observations
should not be obscured by the redshift space provided we use the moment ratio $S_3$ rather than the correlation functions
themselves. However this was only shown to be true for the standard gravitational instability mechanism acting
within the GR framework. The picture could be changed in $f(R)$ gravity. Here the peculiar velocity
power spectra have higher amplitudes at small and moderate scales as shown by \citet{LiHell2013}. This inevitably 
leads to a modified amplitude of the redshift space distortions as shown by \citet{jblkz2012} for the 2-point statistics.
Whether or not this will strongly change the sensitivity to the redshift distortions of the moments ratios $S_n$'s is not known.
This problem must be studied before any galaxy clustering data can be used to constrain $f(R)$ models.

 {\it 3. The effect of baryons.} All our simulations discussed here contain only the DM. The only baryonic effects
we included in the simulations were encoded in the initial transfer functions (the BAO wiggles) and increased 
dark matter density parameter which was set to be equal to the sum of $\Omega_{\rm DM}$ and $\Omega_{\rm b}$. In other words
baryons in our simulations were treated as dark matter. However it is well known that baryon content of the
Universe is a subject of a complicated and intrinsically non-linear hydrodynamical evolution, including effects such as
radiative cooling, reionisation, supernova and AGN feedbacks. It is well established that in the presence of those
effects the higher-order clustering hierarchy can be significantly altered at small scales \citep[\eg][]{skewness_barions}. 
The modified non-linear dynamics of
the $f(R)$ gravity models can only add to the overall complicated picture associated with the baryon content.
Recently \citet{mog_gadget} have used a $f(R)$-enabled version of the \verb#GADGET2# \citep{Gadget2}
to show that at small scales the effects induced by the $f(R)$ gravity in the matter power spectrum 
have the opposite sign to the effects coming from baryonic physics ({\it i.e.} supernova and AGN feedbacks).
The baryonic effects studied by \citet{mog_gadget} appear on similar scales and have comparable magnitudes 
to the effects of the modified $f(R)$ gravity. This shows that there are considerable degeneracies
between the modified gravity and baryonic effects, provided that one is concerned with the power spectrum only.

The general picture emerging from our studies is the following. The $f(R)$ gravity introduces
significant modifications in the higher-order clustering statistics that are especially strong 
at small scales, alas the complicated nature of the galaxy formation process will make it very difficult
to rule out or constrain this class of models using data on spatial clustering of galaxies alone.

\section*{Acknowledgements}
The simulations and their analysis presented in this paper were carried out using the {\it cosmology machine}
at the University of Durham. We would like to acknowledge Lydia Heck whose technical support made our computations
much more smooth. WAH acknowledge the support received from Polish National Science Center in grant no. DEC-2011/01/D/ST9/01960 and ERC Advanced Investigator grant of C.~S.~Frenk, COSMIWAY.
BL is supported by the Royal Astronomical Society and Durham University.
The N-body simulations presented in this paper were run on the ICC Cosmology Machine, which is part of the DiRAC Facility jointly funded by STFC, the Large Facilities Capital Fund of BIS, and Durham University.

\bibliographystyle{mn2e}
\bibliography{clustering_FR}

\begin{thebibliography}{107}
\expandafter\ifx\csname natexlab\endcsname\relax\def\natexlab#1{#1}\fi

\bibitem[{{Amendola}(2000)}]{de5}
{Amendola} L., 2000, \prd, 62, 043511

\bibitem[{{Angulo}, {Baugh} \& {Lacey}(2008){Angulo}, {Baugh}, \&
  {Lacey}}]{npoint-halos}
{Angulo} R.~E., {Baugh} C.~M., {Lacey} C.~G., 2008, \mnras, 387, 921

\bibitem[{{Arag{\'o}n-Calvo} {et~al}\mbox{.}(2010){Arag{\'o}n-Calvo}, {Platen},
  {van de Weygaert}, \& {Szalay}}]{cosmicweb2}
{Arag{\'o}n-Calvo} M.~A., {Platen} E., {van de Weygaert} R., {Szalay} A.~S.,
  2010, \apj, 723, 364

\bibitem[{{Bardeen} {et~al}\mbox{.}(1986){Bardeen}, {Bond}, {Kaiser}, \&
  {Szalay}}]{BBKS}
{Bardeen} J.~M., {Bond} J.~R., {Kaiser} N., {Szalay} A.~S., 1986, \apj, 304, 15

\bibitem[{{Baugh} {et~al}\mbox{.}(2004){Baugh}, {Croton}, {Gazta{\~n}aga},
  {Norberg}, {Colless}, {Baldry}, {Bland-Hawthorn}, {Bridges}, {Cannon},
  {Cole}, {Collins}, {Couch}, {Dalton}, {De Propris}, {Driver}, {Efstathiou},
  {Ellis}, {Frenk}, {Glazebrook}, {Jackson}, {Lahav}, {Lewis}, {Lumsden},
  {Maddox}, {Madgwick}, {Peacock}, {Peterson}, {Sutherland}, {Taylor}, \&
  {2dFGRS Team}}]{2004MNRAS.351L..44B}
{Baugh} C.~M. {et~al.}, 2004, \mnras, 351, L44

\bibitem[{{Baugh}, {Gaztanaga} \& {Efstathiou}(1995){Baugh}, {Gaztanaga}, \&
  {Efstathiou}}]{cic_ana}
{Baugh} C.~M., {Gaztanaga} E., {Efstathiou} G., 1995, \mnras, 274, 1049

\bibitem[{{Bean} {et~al}\mbox{.}(2007){Bean}, {Bernat}, {Pogosian},
  {Silvestri}, \& {Trodden}}]{fr_pertb2}
{Bean} R., {Bernat} D., {Pogosian} L., {Silvestri} A., {Trodden} M., 2007,
  \prd, 75, 064020

\bibitem[{{Bernardeau}(1992)}]{ber1992}
{Bernardeau} F., 1992, \apj, 392, 1

\bibitem[{{Bernardeau}(1994)}]{ber1994}
{Bernardeau} F., 1994, \aap, 291, 697

\bibitem[{{Bernardeau}(2004)}]{Bernardeau_mg}
{Bernardeau} F., 2004, ArXiv Astrophysics e-prints

\bibitem[{{Bernardeau} {et~al}\mbox{.}(2002){Bernardeau}, {Colombi},
  {Gazta{\~n}aga}, \& {Scoccimarro}}]{BCGS_book}
{Bernardeau} F., {Colombi} S., {Gazta{\~n}aga} E., {Scoccimarro} R., 2002,
  \physrep, 367, 1

\bibitem[{{Bond}, {Kofman} \& {Pogosyan}(1996){Bond}, {Kofman}, \&
  {Pogosyan}}]{cosmicweb}
{Bond} J.~R., {Kofman} L., {Pogosyan} D., 1996, \nat, 380, 603

\bibitem[{{Borisov} \& {Jain}(2009)}]{3point_FR}
{Borisov} A., {Jain} B., 2009, \prd, 79, 103506

\bibitem[{{Bouchet} {et~al}\mbox{.}(1995){Bouchet}, {Colombi}, {Hivon}, \&
  {Juszkiewicz}}]{grav_inst_pt2}
{Bouchet} F.~R., {Colombi} S., {Hivon} E., {Juszkiewicz} R., 1995, \aap, 296,
  575

\bibitem[{{Bouchet} \& {Hernquist}(1992)}]{cic_nbody2}
{Bouchet} F.~R., {Hernquist} L., 1992, \apj, 400, 25

\bibitem[{{Brax} {et~al}\mbox{.}(2012){Brax}, {Davis}, {Li}, \&
  {Winther}}]{bdlw2012}
{Brax} P., {Davis} A.-C., {Li} B., {Winther} H.~A., 2012, ArXiv e-prints

\bibitem[{{Brax} {et~al}\mbox{.}(2011){Brax}, {van de Bruck}, {Davis}, {Li}, \&
  {Shaw}}]{bbdls2011}
{Brax} P., {van de Bruck} C., {Davis} A.-C., {Li} B., {Shaw} D.~J., 2011, \prd,
  83, 104026

\bibitem[{{Brax} {et~al}\mbox{.}(2008){Brax}, {van de Bruck}, {Davis}, \&
  {Shaw}}]{bbds2008}
{Brax} P., {van de Bruck} C., {Davis} A.-C., {Shaw} D.~J., 2008, \prd, 78,
  104021

\bibitem[{{Brookfield}, {van de Bruck} \& {Hall}(2006){Brookfield}, {van de
  Bruck}, \& {Hall}}]{bbh2006}
{Brookfield} A.~W., {van de Bruck} C., {Hall} L.~M.~H., 2006, \prd, 74, 064028

\bibitem[{{Carroll}(2001)}]{Carroll2001}
{Carroll} S.~M., 2001, Living Reviews in Relativity, 4, 1

\bibitem[{Carroll {et~al}\mbox{.}(2005)Carroll, De~Felice, Duvvuri, Easson,
  Trodden, {et~al.}}]{cddett2005}
Carroll S.~M., De~Felice A., Duvvuri V., Easson D.~A., Trodden M., {et~al.},
  2005, Phys.Rev., D71, 063513

\bibitem[{{Carroll} {et~al}\mbox{.}(2004){Carroll}, {Duvvuri}, {Trodden}, \&
  {Turner}}]{CDTT2004}
{Carroll} S.~M., {Duvvuri} V., {Trodden} M., {Turner} M.~S., 2004, \prd, 70,
  043528

\bibitem[{{Cautun}, {van de Weygaert} \& {Jones}(2013){Cautun}, {van de
  Weygaert}, \& {Jones}}]{nexus}
{Cautun} M., {van de Weygaert} R., {Jones} B.~J.~T., 2013, \mnras, 429, 1286

\bibitem[{{Cautun} \& {van de Weygaert}(2011)}]{cv2011}
{Cautun} M.~C., {van de Weygaert} R., 2011, ArXiv e-prints

\bibitem[{{Chiba}(2003)}]{Chiba2003}
{Chiba} T., 2003, Physics Letters B, 575, 1

\bibitem[{{Chiba}, {Smith} \& {Erickcek}(2007){Chiba}, {Smith}, \&
  {Erickcek}}]{Chiba2007}
{Chiba} T., {Smith} T.~L., {Erickcek} A.~L., 2007, \prd, 75, 124014

\bibitem[{{Clampitt}, {Cai} \& {Li}(2013){Clampitt}, {Cai}, \&
  {Li}}]{mog_void_exset}
{Clampitt} J., {Cai} Y.-C., {Li} B., 2013, \mnras, 431, 749

\bibitem[{{Cole} {et~al}\mbox{.}(2005){Cole}, {Percival}, {Peacock}, {Norberg},
  {Baugh}, {Frenk}, {Baldry}, {Bland-Hawthorn}, {Bridges}, {Cannon}, {Colless},
  {Collins}, {Couch}, {Cross}, {Dalton}, {Eke}, {De Propris}, {Driver},
  {Efstathiou}, {Ellis}, {Glazebrook}, {Jackson}, {Jenkins}, {Lahav}, {Lewis},
  {Lumsden}, {Maddox}, {Madgwick}, {Peterson}, {Sutherland}, \&
  {Taylor}}]{galaxy_s8_1}
{Cole} S. {et~al.}, 2005, \mnras, 362, 505

\bibitem[{{Colombi}, {Bouchet} \& {Schaeffer}(1994){Colombi}, {Bouchet}, \&
  {Schaeffer}}]{Colombi1994}
{Colombi} S., {Bouchet} F.~R., {Schaeffer} R., 1994, \aap, 281, 301

\bibitem[{{Crocce}, {Pueblas} \& {Scoccimarro}(2006){Crocce}, {Pueblas}, \&
  {Scoccimarro}}]{transients1}
{Crocce} M., {Pueblas} S., {Scoccimarro} R., 2006, \mnras, 373, 369

\bibitem[{{Croton} {et~al}\mbox{.}(2004){Croton}, {Gazta{\~n}aga}, {Baugh},
  {Norberg}, {Colless}, {Baldry}, {Bland-Hawthorn}, {Bridges}, {Cannon},
  {Cole}, {Collins}, {Couch}, {Dalton}, {De Propris}, {Driver}, {Efstathiou},
  {Ellis}, {Frenk}, {Glazebrook}, {Jackson}, {Lahav}, {Lewis}, {Lumsden},
  {Maddox}, {Madgwick}, {Peacock}, {Peterson}, {Sutherland}, \&
  {Taylor}}]{2004MNRAS.352.1232C}
{Croton} D.~J. {et~al.}, 2004, \mnras, 352, 1232

\bibitem[{{Davis} {et~al}\mbox{.}(2012){Davis}, {Li}, {Mota}, \&
  {Winther}}]{dlmw2012}
{Davis} A.-C., {Li} B., {Mota} D.~F., {Winther} H.~A., 2012, \apj, 748, 61

\bibitem[{{de Felice} \& {Tsujikawa}(2010)}]{dt2010}
{de Felice} A., {Tsujikawa} S., 2010, Living Reviews in Relativity, 13, 3

\bibitem[{{Eisenstein} {et~al}\mbox{.}(2005){Eisenstein}, {Zehavi}, {Hogg},
  {Scoccimarro}, {Blanton}, {Nichol}, {Scranton}, {Seo}, {Tegmark}, {Zheng},
  {Anderson}, {Annis}, {Bahcall}, {Brinkmann}, {Burles}, {Castander},
  {Connolly}, {Csabai}, {Doi}, {Fukugita}, {Frieman}, {Glazebrook}, {Gunn},
  {Hendry}, {Hennessy}, {Ivezi{\'c}}, {Kent}, {Knapp}, {Lin}, {Loh}, {Lupton},
  {Margon}, {McKay}, {Meiksin}, {Munn}, {Pope}, {Richmond}, {Schlegel},
  {Schneider}, {Shimasaku}, {Stoughton}, {Strauss}, {SubbaRao}, {Szalay},
  {Szapudi}, {Tucker}, {Yanny}, \& {York}}]{galaxy_s8_2}
{Eisenstein} D.~J. {et~al.}, 2005, \apj, 633, 560

\bibitem[{{Faulkner} {et~al}\mbox{.}(2007){Faulkner}, {Tegmark}, {Bunn}, \&
  {Mao}}]{ftbm2007}
{Faulkner} T., {Tegmark} M., {Bunn} E.~F., {Mao} Y., 2007, \prd, 76, 063505

\bibitem[{{Feldman} {et~al}\mbox{.}(2003){Feldman}, {Juszkiewicz}, {Ferreira},
  {Davis}, {Gazta{\~n}aga}, {Fry}, {Jaffe}, {Chambers}, {da Costa}, {Bernardi},
  {Giovanelli}, {Haynes}, \& {Wegner}}]{Om_sigma8_pv}
{Feldman} H. {et~al.}, 2003, \apjl, 596, L131

\bibitem[{{Fry}(1984{\natexlab{a}})}]{Fry1984b}
{Fry} J.~N., 1984{\natexlab{a}}, \apjl, 277, L5

\bibitem[{{Fry}(1984{\natexlab{b}})}]{Fry1984a}
{Fry} J.~N., 1984{\natexlab{b}}, \apj, 279, 499

\bibitem[{{Fry} \& {Gaztanaga}(1993)}]{FG1993}
{Fry} J.~N., {Gaztanaga} E., 1993, \apj, 413, 447

\bibitem[{{Fry} \& {Gaztanaga}(1994)}]{FG1994}
{Fry} J.~N., {Gaztanaga} E., 1994, \apj, 425, 1

\bibitem[{{Gazta{\~n}aga} {et~al}\mbox{.}(2005){Gazta{\~n}aga}, {Norberg},
  {Baugh}, \& {Croton}}]{2005MNRAS.364..620G}
{Gazta{\~n}aga} E., {Norberg} P., {Baugh} C.~M., {Croton} D.~J., 2005, \mnras,
  364, 620

\bibitem[{{Gaztanaga}(1994)}]{GaztanagaAPM94}
{Gaztanaga} E., 1994, \mnras, 268, 913

\bibitem[{{Gaztanaga} \& {Baugh}(1995)}]{cic_nbody1}
{Gaztanaga} E., {Baugh} C.~M., 1995, \mnras, 273, L1

\bibitem[{{Gaztanaga} \& {Bernardeau}(1998)}]{grav_inst_pt3}
{Gaztanaga} E., {Bernardeau} F., 1998, \aap, 331, 829

\bibitem[{{Goroff} {et~al}\mbox{.}(1986){Goroff}, {Grinstein}, {Rey}, \&
  {Wise}}]{smoothing_comute}
{Goroff} M.~H., {Grinstein} B., {Rey} S.-J., {Wise} M.~B., 1986, \apj, 311, 6

\bibitem[{{Guillet}, {Teyssier} \& {Colombi}(2009){Guillet}, {Teyssier}, \&
  {Colombi}}]{skewness_barions}
{Guillet} T., {Teyssier} R., {Colombi} S., 2009, ArXiv e-prints

\bibitem[{{Hellwing}(2013)}]{biasFR}
{Hellwing}, W. {\it et al}., 2013, in preparation

\bibitem[{{Hellwing} \& {Juszkiewicz}(2009)}]{rebel1}
{Hellwing} W.~A., {Juszkiewicz} R., 2009, Phys.~Rev.~D, 80, 083522

\bibitem[{{Hellwing}, {Juszkiewicz} \& {van de Weygaert}(2010){Hellwing},
  {Juszkiewicz}, \& {van de Weygaert}}]{Hellwing_npoint}
{Hellwing} W.~A., {Juszkiewicz} R., {van de Weygaert} R., 2010, \prd, 82,
  103536

\bibitem[{{Hellwing} {et~al}\mbox{.}(2013){Hellwing}, {Juszkiewicz}, {van de
  Weygaert}, \& {Bilicki}}]{SkewBAO2}
{Hellwing} W.~A., {Juszkiewicz} R., {van de Weygaert} R., {Bilicki} M., 2013,
  ArXiv e-prints

\bibitem[{{Hinshaw} {et~al}\mbox{.}(2012){Hinshaw}, {Larson}, {Komatsu},
  {Spergel}, {Bennett}, {Dunkley}, {Nolta}, {Halpern}, {Hill}, {Odegard},
  {Page}, {Smith}, {Weiland}, {Gold}, {Jarosik}, {Kogut}, {Limon}, {Meyer},
  {Tucker}, {Wollack}, \& {Wright}}]{WMAP9}
{Hinshaw} G. {et~al.}, 2012, ArXiv e-prints

\bibitem[{{Hivon} {et~al}\mbox{.}(1995){Hivon}, {Bouchet}, {Colombi}, \&
  {Juszkiewicz}}]{grav_inst_pt4}
{Hivon} E., {Bouchet} F.~R., {Colombi} S., {Juszkiewicz} R., 1995, \aap, 298,
  643

\bibitem[{{Hu} \& {Sawicki}(2007)}]{hs2007}
{Hu} W., {Sawicki} I., 2007, \prd, 76, 064004

\bibitem[{{Jennings} {et~al}\mbox{.}(2012){Jennings}, {Baugh}, {Li}, {Zhao}, \&
  {Koyama}}]{jblkz2012}
{Jennings} E., {Baugh} C.~M., {Li} B., {Zhao} G.-B., {Koyama} K., 2012, ArXiv
  e-prints

\bibitem[{{Jennings}, {Li} \& {Hu}(2013){Jennings}, {Li}, \&
  {Hu}}]{JenningsVoid}
{Jennings} E., {Li} Y., {Hu} W., 2013, ArXiv e-prints

\bibitem[{{Juszkiewicz} \& {Bouchet}(1996)}]{cl_stat_dyn}
{Juszkiewicz} R., {Bouchet} F.~R., 1996, ArXiv Astrophysics e-prints

\bibitem[{{Juszkiewicz}, {Bouchet} \& {Colombi}(1993){Juszkiewicz}, {Bouchet},
  \& {Colombi}}]{Juszkiewicz1993}
{Juszkiewicz} R., {Bouchet} F.~R., {Colombi} S., 1993, \apj, 412, L9

\bibitem[{{Juszkiewicz} {et~al}\mbox{.}(2010){Juszkiewicz}, {Feldman}, {Fry},
  \& {Jaffe}}]{RomanSimga8}
{Juszkiewicz} R., {Feldman} H.~A., {Fry} J.~N., {Jaffe} A.~H., 2010, \jcap, 2,
  21

\bibitem[{{Juszkiewicz}, {Hellwing} \& {van de Weygaert}(2013){Juszkiewicz},
  {Hellwing}, \& {van de Weygaert}}]{SkewBAO}
{Juszkiewicz} R., {Hellwing} W.~A., {van de Weygaert} R., 2013, \mnras, 429,
  1206

\bibitem[{{Kamenshchik}, {Moschella} \& {Pasquier}(2001){Kamenshchik},
  {Moschella}, \& {Pasquier}}]{de4}
{Kamenshchik} A., {Moschella} U., {Pasquier} V., 2001, Physics Letters B, 511,
  265

\bibitem[{{Kendall} \& {Stuart}(1977)}]{1977KendallStuart}
{Kendall} M., {Stuart} A., 1977, {The advanced theory of statistics. Vol.1:
  Distribution theory}

\bibitem[{{Khoury} \& {Weltman}(2004)}]{kw2004}
{Khoury} J., {Weltman} A., 2004, \prd, 69, 044026

\bibitem[{{Koivisto}(2006)}]{matter_pk_FR}
{Koivisto} T., 2006, \prd, 73, 083517

\bibitem[{{Li}(2011)}]{scalar_voids_Li}
{Li} B., 2011, \mnras, 411, 2615

\bibitem[{{Li} \& {Barrow}(2007)}]{lb2007}
{Li} B., {Barrow} J.~D., 2007, \prd, 75, 084010

\bibitem[{{Li} \& {Barrow}(2011)}]{lb2011}
{Li} B., {Barrow} J.~D., 2011, \prd, 83, 024007

\bibitem[{{Li} {et~al}\mbox{.}(2013){Li}, {Hellwing}, {Koyama}, {Zhao},
  {Jennings}, \& {Baugh}}]{LiHell2013}
{Li} B., {Hellwing} W.~A., {Koyama} K., {Zhao} G.-B., {Jennings} E., {Baugh}
  C.~M., 2013, \mnras, 428, 743

\bibitem[{{Li}, {Zhao} \& {Koyama}(2012){Li}, {Zhao}, \&
  {Koyama}}]{halo_void_fr}
{Li} B., {Zhao} G.-B., {Koyama} K., 2012, \mnras, 421, 3481

\bibitem[{{Li} {et~al}\mbox{.}(2012){Li}, {Zhao}, {Teyssier}, \&
  {Koyama}}]{lztk2012}
{Li} B., {Zhao} G.-B., {Teyssier} R., {Koyama} K., 2012, \jcap, 1, 51

\bibitem[{{Li} \& {Zhao}(2009)}]{lz2009}
{Li} B., {Zhao} H., 2009, \prd, 80, 044027

\bibitem[{{Li} \& {Zhao}(2010)}]{lz2010}
{Li} B., {Zhao} H., 2010, \prd, 81, 104047

\bibitem[{{\L okas} {et~al}\mbox{.}(1995){\L okas}, {Juszkiewicz}, {Weinberg},
  \& {Bouchet}}]{Lokas_kurt}
{\L okas} E.~L., {Juszkiewicz} R., {Weinberg} D.~H., {Bouchet} F.~R., 1995,
  \mnras, 274, 730

\bibitem[{{Mota} \& {Shaw}(2007)}]{ms2007}
{Mota} D.~F., {Shaw} D.~J., 2007, \prd, 75, 063501

\bibitem[{{Navarro} \& {Van Acoleyen}(2007)}]{nv2007}
{Navarro} I., {Van Acoleyen} K., 2007, \jcap, 2, 22

\bibitem[{{Oyaizu}(2008)}]{oyaizu2008}
{Oyaizu} H., 2008, \prd, 78, 123523

\bibitem[{{Oyaizu}, {Lima} \& {Hu}(2008){Oyaizu}, {Lima}, \& {Hu}}]{olh2008}
{Oyaizu} H., {Lima} M., {Hu} W., 2008, \prd, 78, 123524

\bibitem[{{Paranjape}, {Lam} \& {Sheth}(2012){Paranjape}, {Lam}, \&
  {Sheth}}]{voidado2}
{Paranjape} A., {Lam} T.~Y., {Sheth} R.~K., 2012, \mnras, 420, 1648

\bibitem[{{Peebles}(1980)}]{1980Peebles}
{Peebles} P.~J.~E., 1980, {The large-scale structure of the universe}. Research
  supported by the National Science Foundation.~Princeton, N.J., Princeton
  University Press, 1980.~435 p.

\bibitem[{{Peebles} \& {Ratra}(1988)}]{de2}
{Peebles} P.~J.~E., {Ratra} B., 1988, \apjl, 325, L17

\bibitem[{{Perlmutter} {et~al}\mbox{.}(1999){Perlmutter}, {Aldering},
  {Goldhaber}, {Knop}, {Nugent}, {Castro}, {Deustua}, {Fabbro}, {Goobar},
  {Groom}, {Hook}, {Kim}, {Kim}, {Lee}, {Nunes}, {Pain}, {Pennypacker},
  {Quimby}, {Lidman}, {Ellis}, {Irwin}, {McMahon}, {Ruiz-Lapuente}, {Walton},
  {Schaefer}, {Boyle}, {Filippenko}, {Matheson}, {Fruchter}, {Panagia},
  {Newberg}, {Couch}, \& {The Supernova Cosmology Project}}]{acceleration2}
{Perlmutter} S. {et~al.}, 1999, \apj, 517, 565

\bibitem[{{Planck Collaboration} {et~al}\mbox{.}(2013){Planck Collaboration},
  {Ade}, {Aghanim}, {Armitage-Caplan}, {Arnaud}, {Ashdown}, {Atrio-Barandela},
  {Aumont}, {Baccigalupi}, {Banday}, \& et~al.}]{Planck1}
{Planck Collaboration} {et~al.}, 2013, ArXiv e-prints

\bibitem[{{Puchwein}, {Baldi} \& {Springel}(2013){Puchwein}, {Baldi}, \&
  {Springel}}]{mog_gadget}
{Puchwein} E., {Baldi} M., {Springel} V., 2013, ArXiv e-prints

\bibitem[{{Ratra} \& {Peebles}(1988)}]{de1}
{Ratra} B., {Peebles} P.~J.~E., 1988, \prd, 37, 3406

\bibitem[{{Riess} {et~al}\mbox{.}(1998){Riess}, {Filippenko}, {Challis},
  {Clocchiatti}, {Diercks}, {Garnavich}, {Gilliland}, {Hogan}, {Jha},
  {Kirshner}, {Leibundgut}, {Phillips}, {Reiss}, {Schmidt}, {Schommer},
  {Smith}, {Spyromilio}, {Stubbs}, {Suntzeff}, \& {Tonry}}]{acceleration1}
{Riess} A.~G. {et~al.}, 1998, \aj, 116, 1009

\bibitem[{{Ross}, {Brunner} \& {Myers}(2007){Ross}, {Brunner}, \&
  {Myers}}]{skewness_obs1}
{Ross} A.~J., {Brunner} R.~J., {Myers} A.~D., 2007, \apj, 665, 67

\bibitem[{{Schaap} \& {van de Weygaert}(2000)}]{sv2000}
{Schaap} W.~E., {van de Weygaert} R., 2000, \aap, 363, L29

\bibitem[{{Schmidt}(2009)}]{schmidt2009}
{Schmidt} F., 2009, \prd, 80, 043001

\bibitem[{{Schmidt} {et~al}\mbox{.}(2009){Schmidt}, {Lima}, {Oyaizu}, \&
  {Hu}}]{sloh2009}
{Schmidt} F., {Lima} M., {Oyaizu} H., {Hu} W., 2009, \prd, 79, 083518

\bibitem[{Schmidt, Vikhlinin \& Hu(2009)Schmidt, Vikhlinin, \& Hu}]{svh2009}
Schmidt F., Vikhlinin A., Hu W., 2009, Phys.Rev., D80, 083505

\bibitem[{{Scoccimarro}(1998)}]{transients3}
{Scoccimarro} R., 1998, \mnras, 299, 1097

\bibitem[{{Shandarin}, {Habib} \& {Heitmann}(2012){Shandarin}, {Habib}, \&
  {Heitmann}}]{cosmicweb3}
{Shandarin} S., {Habib} S., {Heitmann} K., 2012, \prd, 85, 083005

\bibitem[{{Sheth} \& {van de Weygaert}(2004)}]{voidado1}
{Sheth} R.~K., {van de Weygaert} R., 2004, \mnras, 350, 517

\bibitem[{{Song}, {Hu} \& {Sawicki}(2007){Song}, {Hu}, \&
  {Sawicki}}]{fr_pertb1}
{Song} Y.-S., {Hu} W., {Sawicki} I., 2007, \prd, 75, 044004

\bibitem[{{Sotiriou} \& {Faraoni}(2010)}]{sf2010}
{Sotiriou} T.~P., {Faraoni} V., 2010, Reviews of Modern Physics, 82, 451

\bibitem[{{Springel}(2005)}]{Gadget2}
{Springel} V., 2005, Mon.~Not.~Roy.~Astron.~Soc., 364, 1105

\bibitem[{{Szapudi} \& {Colombi}(1996)}]{cosmic_error}
{Szapudi} I., {Colombi} S., 1996, \apj, 470, 131

\bibitem[{{Szapudi} {et~al}\mbox{.}(1999){Szapudi}, {Quinn}, {Stadel}, \&
  {Lake}}]{npoint_omega_cdm}
{Szapudi} I., {Quinn} T., {Stadel} J., {Lake} G., 1999, \apj, 517, 54

\bibitem[{{Tatekawa} \& {Mizuno}(2007)}]{transients2}
{Tatekawa} T., {Mizuno} S., 2007, Journal of Cosmology and Astro-Particle
  Physics, 12, 14

\bibitem[{{Tatekawa} \& {Tsujikawa}(2008)}]{skewness_FR}
{Tatekawa} T., {Tsujikawa} S., 2008, \jcap, 9, 9

\bibitem[{{Tegmark} {et~al}\mbox{.}(2004){Tegmark}, {Blanton}, {Strauss},
  {Hoyle}, {Schlegel}, {Scoccimarro}, {Vogeley}, {Weinberg}, {Zehavi},
  {Berlind}, {Budavari}, {Connolly}, {Eisenstein}, {Finkbeiner}, {Frieman},
  {Gunn}, {Hamilton}, {Hui}, {Jain}, {Johnston}, {Kent}, {Lin}, {Nakajima},
  {Nichol}, {Ostriker}, {Pope}, {Scranton}, {Seljak}, {Sheth}, {Stebbins},
  {Szalay}, {Szapudi}, {Verde}, {Xu}, {Annis}, {Bahcall}, {Brinkmann},
  {Burles}, {Castander}, {Csabai}, {Loveday}, {Doi}, {Fukugita}, {Gott},
  {Hennessy}, {Hogg}, {Ivezi{\'c}}, {Knapp}, {Lamb}, {Lee}, {Lupton}, {McKay},
  {Kunszt}, {Munn}, {O'Connell}, {Peoples}, {Pier}, {Richmond}, {Rockosi},
  {Schneider}, {Stoughton}, {Tucker}, {Vanden Berk}, {Yanny}, {York}, \& {SDSS
  Collaboration}}]{galaxy_s8_3}
{Tegmark} M. {et~al.}, 2004, \apj, 606, 702

\bibitem[{{Teyssier}(2002)}]{ramses}
{Teyssier} R., 2002, \aap, 385, 337

\bibitem[{{van de Weygaert} \& {Schaap}(2009)}]{vs2009}
{van de Weygaert} R., {Schaap} W., 2009, in Lecture Notes in Physics, Berlin
  Springer Verlag, Vol. 665, Data Analysis in Cosmology, {Mart{\'{\i}}nez}
  V.~J., {Saar} E., {Mart{\'{\i}}nez-Gonz{\'a}lez} E., {Pons-Border{\'{\i}}a}
  M.-J., eds., pp. 291--413

\bibitem[{{Watkins}, {Feldman} \& {Hudson}(2009){Watkins}, {Feldman}, \&
  {Hudson}}]{s8_bulk_flow}
{Watkins} R., {Feldman} H.~A., {Hudson} M.~J., 2009, \mnras, 392, 743

\bibitem[{{Zaldarriaga}, {Seljak} \& {Hui}(2001){Zaldarriaga}, {Seljak}, \&
  {Hui}}]{2001ApJ...551...48Z}
{Zaldarriaga} M., {Seljak} U., {Hui} L., 2001, \apj, 551, 48

\bibitem[{{Zel'Dovich}(1970)}]{za}
{Zel'Dovich} Y.~B., 1970, \aap, 5, 84

\bibitem[{{Zhao}, {Li} \& {Koyama}(2011){Zhao}, {Li}, \& {Koyama}}]{zlk2011}
{Zhao} G.-B., {Li} B., {Koyama} K., 2011, \prd, 83, 044007

\bibitem[{{Zlatev}, {Wang} \& {Steinhardt}(1999){Zlatev}, {Wang}, \&
  {Steinhardt}}]{de3}
{Zlatev} I., {Wang} L., {Steinhardt} P.~J., 1999, Physical Review Letters, 82,
  896

\end{thebibliography}

\bsp

\label{lastpage}

\end{document}